\documentclass[11pt,oneside,letterpaper]{article}
\usepackage{amssymb}
\usepackage{amsmath}
\usepackage[dvips]{graphicx}
\usepackage{setspace}
\usepackage{xcolor}
\usepackage{ifpdf}
\usepackage{graphicx}

\newcommand{\p}{\partial}
\newcommand{\bx}{\textbf{X}}

\numberwithin{equation}{section}

\def\half{{1\over 2}}

\def\e{{\epsilon}}
\def\p{\partial}

\newcommand{\bi}{\begin{itemize}}
\newcommand{\ei}{\end{itemize}}
\newcommand{\bea}{\begin{eqnarray}}
\newcommand{\eea}{\end{eqnarray}}
\newcommand{\be}{\begin{equation}}
\newcommand{\ee}{\end{equation}}
\newcommand{\non}{\nonumber}

\allowdisplaybreaks  

\def\C{\Bbb{C}}
\def\Z{\Bbb{Z}}

\def\R{\Bbb{R}}
\def\H{\Bbb{H}}

\def\rref#1{(\ref{#1})}

\begin{document}
\vspace*{2.5cm}
\begin{center}
{ \LARGE \textbf{Chiral Gravity, Log Gravity and Extremal CFT }\\}
\vskip 1.5cm
{Alexander Maloney$^{\dag}$, Wei Song$^{\ddag\S}$ and
Andrew Strominger$^\S$}
 \vskip 0.9 cm

{\em \small $^\dag$ Physics Department, McGill University,
MontrÁŠeal, QC H3A 2T8, Canada}
 \vskip 0.1 cm
{\em \small $^\ddag$ Key Laboratory of Frontiers in Theoretical
Physics,\\ Institute of Theoretical Physics, Chinese Academy of
Sciences, 
Beijing, 100190, China}
 \vskip 0.1 cm
{\em \small $^\S$Center for the Fundamental Laws of Nature\\
Jefferson Physical Laboratory, Harvard University, Cambridge, MA 02138, USA}\\


\end{center}
\begin{abstract}

We show that the linearization of all exact solutions of
classical chiral gravity around the AdS$_3$ vacuum have positive
energy.  Non-chiral and negative-energy solutions of the linearized
equations are infrared divergent at second order, and so are removed from the spectrum.  In other words, chirality is confined and the equations of motion have linearization instabilities.
We prove that the only stationary, axially symmetric solutions of chiral gravity are BTZ black holes, which have positive energy. It is further shown that classical log gravity
-- the theory with logarithmically relaxed boundary conditions --
has finite asymptotic symmetry generators but is not chiral and  hence may be dual at the quantum level to a logarithmic CFT.   Moreover we  show that log gravity contains
chiral gravity within it as a decoupled charge superselection sector. We formally evaluate
the Euclidean sum over geometries of chiral gravity and show that it gives precisely the holomorphic extremal CFT partition function. The modular
invariance and integrality of the expansion coefficients of this partition function are consistent with the existence of an exact quantum theory
of chiral gravity.  We argue that the problem of quantizing
chiral gravity is the holographic dual of the problem of
constructing an extremal CFT, while quantizing log gravity is  dual
to the problem of constructing a logarithmic extremal CFT.

\noindent \end{abstract}

\newpage
\setcounter{page}{1}
\pagenumbering{arabic}

\tableofcontents

\onehalfspacing


\section{Introduction}
A consistent, non-trivial theory of pure gravity in three dimensions
-- classical or quantum -- with a stable vacuum would undoubtedly
provide invaluable insights  into the many complexities of gravity
in our four-dimensional world.  Unfortunately, pure 3D Einstein
gravity is locally trivial classically, while its quantum status
remains unclear despite decades of investigations.  Recently,  an
exceptional and clearly nontrivial 3D theory termed ``chiral
gravity" was discovered \cite{lss}. This theory is a special case of topologically massive gravity \cite{Deser:1981wh,Deser:1982vy}
at a particular value of the couplings, and is defined with  asymptotically
AdS$_3$ boundary conditions, in the sense of Fefferman-Graham-Brown-Henneaux \cite{fg, Brown:1986nw}.\footnote{Chiral gravity differs in this respect  from  log gravity which has the same action but logarithmically weaker boundary conditions.} It was conjectured in \cite{lss} that at the classical level
\begin{itemize}
\item
Chiral gravity is chiral, in the sense that the asymptotic symmetry
group is generated by a single copy of the Virasoro algebra,
\footnote{The quantum version of this conjecture is that physical
states form representations of a single Virasoro algebra.}
\item
Solutions of chiral gravity have positive energy.
\end{itemize}
Some supporting evidence was given \cite{lss}. Should both conjectures turn
out to be true, chiral gravity, in its quantum version,  would prove an extremely
interesting gedanken laboratory for the study of quantum gravity.

The chirality and positivity conjectures generated some controversy.
Shortly after \cite{lss}, interesting new solutions to the linearized
equations which are not global energy eigenmodes and have a variety
of asymptotic behaviors were discovered. These solutions are
non-chiral and/or negative-energy and were argued to provide
counterexamples to one or both of the classical conjectures Ê\cite{
Carlip:2008jk, Giribet:2008bw} -- see also \cite{
Grumiller:2008qz,Park:2008yy, Grumiller:2008pr, cdww, Carlip:2008qh}.
Subsequently the chirality conjecture was proven \cite{as} and the
claims that these modes provide counterexamples to this conjecture
were revised or withdrawn \cite{Carlip:2008jk2,Giribet:2008bw2}. ÊA
proof of the chirality
conjecture in a different formalism appeared in
\cite{Carlip:2008qh2}. ÊNevertheless, claims that the proposed counterexamples
disprove the positivity conjecture remain in the literature. However
the modes exhibited in
\cite{Carlip:2008jk,Giribet:2008bw} explicitly
violate the
chirality conjecture as well as the positivity conjecture. So if these
modes are truly present in the linearization of the exact spectrum,
they are fatally at odds not only with positivity but with the
chirality proofs of \cite{as, Carlip:2008qh2}. In short, the
literature contains contradictory claims. For related work, see
\cite{Li:2008yz,Sachs:2008gt,Lee:2008gta,Sachs:2008yi,Gibbons:2008vi,Myung:2008dm,Alishahiha:2008rt,Kim:2008bz,Sezgin:2009dj}.

In this paper, we reconcile all these computations and hope to
thereby resolve the controversy. In the process, a perturbative
version of the positivity conjecture will be established to first
order in the deviation around AdS$_3$. The alleged counterexamples
do not disprove positivity for exactly the same reason that they do
not disprove chirality: the equations have a linearization
instability. At second order in perturbation theory, explicit
computation reveals that the metric perturbation develops an
infrared singularity, growing logarithmically with the radius at
infinity. Hence these solutions of the linearized equations are
$not$ the linear approximation to any exact solution of the theory
with Brown-Henneaux boundary conditions.  In other words, chirality
is confined and chiral gravity has linearization instabilities.\footnote{Similar linearization instabilities have occurred in a
number of contexts in general relativity, see e.g. \cite{desa,
ger,bv}.} This divergence was bound to appear because otherwise
there would be a discrepancy between the surface integral expression
for the energy (linear in the second order perturbation) and the
bulk expression (quadratic in the first order perturbation).  The
first of these is manifestly chiral for asymptotically AdS$_3$
solutions, while the second gives a non-chiral answer.  A key
ingredient in reconciling herein the various computations is the
discovery (independently made in \cite{hmt}) of previously neglected
terms in the boundary expressions for the Virasoro charges. The
omission of these terms has led to some contradictory statements in
the literature.

An (imperfect) analogy can be found in QCD.  In the linearized approximation,
the theory contains free quarks. But there is an infrared divergence
in the back reaction caused by the quark and the exact finite energy
spectrum contains only color singlets. A free quark is not a valid
linearized approximation to any finite energy QCD state. Of
course, if the boundary conditions are relaxed to allow flux tubes at
infinity there are single-quark solutions. We will see below that an
analogous relaxation of the boundary conditions for chiral gravity
to those of log gravity allows for non-chiral excitations with finite charges.

The analogy here is imperfect in that color confinement in QCD is a
difficult non-perturbative problem. In contrast, confinement in
chiral gravity can be seen explicitly in second order perturbation
theory. Moreover, in QCD color confinement gives one global
constraint, while in chiral gravity there are an infinite number of
constraints arising from the infinity of conserved (left) Virasoro
charges, all of which must vanish. This is exactly what is required
to eliminate an entire chiral half of the spectrum, and reconcile
the chiral nature of the theory with the non-chiral ``bulk" degree of
freedom found in the local analyses of \cite{Carlip:2008jk,
Grumiller:2008qz,Park:2008yy,Grumiller:2008pr, cdww, Carlip:2008qh2,
Giribet:2008bw,Blagojevic:2008bn}. Rather, we will see below these
local analyses apply to log gravity.

The miraculous escape of chiral gravity from the alleged
perturbative instabilities leads one to hope that there is an exact
positive energy theorem for the theory. The proof of such a theorem
at the non-perturbative level remains an outstanding challenge. We
take one step in this direction by proving a Birkhoff-like theorem:
all stationary, axially symmetric solutions of chiral gravity are
BTZ black holes. The difficulty we encounter in what would seem a
straightforward exercise illustrates the complexity of the full
nonlinear equations. It is interesting to note that all known
solutions of chiral gravity are also solutions of the Einstein
equation. This may be the case for all solutions, although we will
not attempt to demonstrate this here.  One might also attempt to
prove a version of cosmic censorship for chiral gravity.

Armed with knowledge the perturbative spectrum, we then move
on to an analysis of the quantum problem. We apply the standard
methodology of Euclidean quantum gravity to compute the torus
partition function as a function of the modular parameter $\tau$.
Euclidean quantum gravity is, for a variety of reasons,  a
notoriously treacherous subject and the present application
cannot be regarded as completely rigorous. Nevertheless the results are highly
encouraging.  We show that all real saddle points solve the Einstein
equation, and can be classified. Moreover, at the chiral
value of the coupling constants the Euclidean action is
holomorphic.  Following \cite{Maloney:2007ud},
we perform the sum over saddle points including all perturbative corrections, formally obtaining the exact
answer for the partition function. The result is
simply the ``chiral part" of the extremal partition function
conjectured by Witten \cite{Witten:2007kt} to be dual to 3D Einstein
gravity. It is invariant under modular transformations
and has an expansion in $q=e^{2\pi i \tau}$ with integer
coefficients, as required for a consistent quantum mechanical
interpretation as a Hilbert space trace. The spectrum reproduces the
entropy of the BTZ black hole, including both the Bekenstein-Hawking
piece and an infinite series of corrections.  Although it is not
known whether a CFT exists which realizes this spectrum, the encouraging outcome of this computation
might be regarded as evidence both for the existence of quantum chiral gravity
as well as for the existence of such CFTs. In any case  the
interesting problems of understanding quantum chiral gravity and
extremal CFTs are clearly closely linked.

We also consider the theory of log gravity introduced in
\cite{Grumiller:2008qz,Grumiller:2008es}. This theory has the same
action as chiral gravity, but the boundary conditions are weakened
to allow metric fluctuations which grow logarithmically with the
proper radius. Log gravity contains a rich and interesting class of
solutions \cite{AyonBeato:2004fq, cdww, Gibbons:2008vi,
Garbarz:2008qn} which are excluded in chiral gravity. In particular,
the linearization of the exact spectrum includes the non-chiral
modes of \cite{ Carlip:2008jk, Giribet:2008bw, Grumiller:2008qz},
which appear in indecomposable Virasoro representations. The relaxed
boundary conditions also lead to zero-norm states, violations of
unitary and violations of positivity. Interestingly, these
violations resemble those found in logarithmic CFTs, suggesting that
log gravity is dual to a logarithmic CFT \cite{Grumiller:2008qz}.
We show here that the log gravity boundary conditions lead to finite
expressions for the asymptotic symmetry generators.  However,
contrary to \cite{Grumiller:2008es}, the generators are $not$
chiral.  This is consistent with the conjecture that log gravity is
dual to a logarithmic CFT, as logarithmic conformal field theories
cannot be chiral.  We also show that log gravity contains within it
chiral gravity as the superselection sector with vanishing left
Virasoro charges. Thus although log gravity itself is not unitary,
it has a potentially unitary ``physical subspace". We speculate
herein that log gravity may be dual to an "extremal"  logarithmic
CFT whose partition function coincides with Witten's extremal
partition function.

This paper is organized as follows. Section 2 contains basic
formulae and conventions. In section 3 we give the new expression
for the asymptotic symmetry generators. In section 4 we work out the
perturbation expansion around AdS$_3$ to second order, and show
that the non-chiral negative energy solutions to the linearized
equations are not the linearization of exact solutions. In section 5 we study the spectrum at the non-linear level, and prove a Birkhoff-like theorem for stationary, axially symmetric solutions.  In section 6 we study log gravity, show that the asymptotic symmetry group has finite generators and discuss the problem of constructing a symplectic form as required for a canonical formulation.  We show that although log gravity is non-chiral, it contains chiral gravity as a superselection sector. In section 7 we evaluate the
Euclidean partition function and show that it gives the modular
invariant extremal partition function. Finally section 8 concludes
with a discussion of and speculations on the fascinating relation between chiral gravity, log
gravity, extremal CFT and extremal logarithmic CFT.

As this work was nearing completion, the eprint \cite{hmt} appeared with results which
overlap with sections 3 and 6.1. All points in common are in precise agreement.

\section{Preliminaries}
In this section we record some pertinent formulae and establish notation.
Chiral gravity is a special case of topologically massive gravity
(TMG) \cite{Deser:1981wh,Deser:1982vy} with a negative cosmological
constant. TMG is described by the action \be\label{action}
I_{TMG} = \frac{1}{16\pi G}\left[\int d^3x\sqrt{-g}(R+2/\ell^2)+{1
\over \mu}I_{CS}\right] \ee where $I_{CS}$ is the gravitational
Chern-Simons action \be\label{csaction} I_{CS} = \frac{1}{2 }\int_{\mathcal{M}}
d^3x\sqrt{-g}
\varepsilon^ {\lambda\mu\nu}\Gamma^{r}_{\lambda
\sigma}\left(\partial_{\mu}\Gamma^ \sigma_{{r}
\nu}+\frac{2}{3}\Gamma^\sigma_{\mu\tau}\Gamma^\tau_{\nu {r}} \right)
\ee
 and $G$ has the conventional positive sign. The equation of motion in TMG is \be\label{eom}E_{\mu\nu}\equiv
\mathcal{G}_{\mu\nu}+{1\over\mu}C_{\mu\nu}=0 ,\ee where we have
defined \be\label{cgdef} C_{\mu\nu}\equiv \epsilon^{\alpha\beta}\,_{(\mu}
\mathcal{G}_{\nu)\beta;\alpha},~~~~~
\mathcal{G}_{\mu\nu}\equiv{G}_{\mu\nu}-{1 \over
\ell^2}g_{\mu\nu}.\ee These equations have the vacuum solution
\bea\label{gads}
ds^2&=&\ell^2\left(-\cosh^2\rho d\tau^2+\sinh^2\rho d\phi^2 +d\rho^2\right)\non\\
 &=&{\ell^2\over4}\left(-2{\cosh2\rho} d\tau^+d\tau^--d\tau^{+2}-d\tau^{-2}+4d\rho^2\right),\\
 &~&~~~~~~~~~~\tau^\pm={\tau \pm \phi}.\non
\eea
 Chiral gravity \cite{lss} is defined by taking $\mu\ell \to 1$ while keeping the standard Brown-Henneaux \cite{Brown:1986nw} asymptotically AdS$_3$ boundary conditions.
These require that fluctuations $h_{\mu\nu}$ of the metric about
(\ref{gads}) fall off at the boundary according to
  \be\label{strictbc}
\left(
  \begin{array}{ccccc}
 h_{++}= \mathcal{O}({ 1}) & h_{+-}= \mathcal{O}({1})  &h_{+\rho}= \mathcal{O}({e^{-2\rho}})  \\
 h_{-+}=h_{+-} & h_{--}= \mathcal{O}({1})  &h_{-\rho}= \mathcal{O}({e^{-2\rho}})  \\   h_{\rho+}=h_{+\rho} & h_{\rho-}=h_{-\rho} & h_{\rho\rho}= \mathcal{O}({e^{-2\rho}}) \\
  \end{array}
\right) \ee The consistency of these boundary conditions for all
values of $\mu$ was demonstrated in \cite{Hotta:2008yq}. The most
general diffeomorphism which preserves
 (\ref{strictbc}) is of the form
\bea\label{bcp}
\zeta&=&\zeta^+\p_++\zeta^-\p_-+\zeta^\rho\p_\rho\\ \non
&=&\left[ \epsilon^+(\tau^+) +{ 2 e^{-2\rho}}\p_-^2\epsilon^-(\tau^-)+
\mathcal{O}(e^{-4\rho})\right]\p_+
\\ &~~+&\left[ \epsilon^-(\tau^-) +{2e^{-2\rho}}\p_+^2\epsilon^+(\tau^+)+
\mathcal{O}(e^{-4\rho})\right]\p_-\non\\ &~~-&\half \left[
\p_+\epsilon^+(\tau^+) +\p_-\epsilon^-(\tau^-)+
\mathcal{O}(e^{-2\rho})\right]\p_\rho. \non\eea These are parameterized by
a left moving function $\epsilon^-(\tau^-)$ and a right moving function
$\epsilon^+(\tau^+)$. We denote diffeomorphisms depending only on $\e^-$ by
$\xi^L$ and those depending only on $\e^+$ as $\xi^R$. The
subleading terms all correspond to trivial diffeomorphisms;
their generators have no surface term and hence vanish when the
constraints are imposed.  The asymptotic symmetry group (ASG) is
defined as the general boundary-condition-preserving diffeomorphism
(\ref{bcp}) modulo the trivial diffeomorphisms. For generic $\mu$
the ASG is generated by two copies of the Virasoro algebra, which
may be taken to be \be \xi^L_n=\xi(\e^-=e^{in(\tau-\phi)},
\e^+=0)~~~~ \xi^R_n=\xi(\e^-=0,\e^+=e^{in(\tau+\phi)}).\ee  These of
course have a global $SL(2,\mathbb{R})_L\times SL(2,\mathbb{R})_R$
subgroup which generates the AdS$_3$ isometries. At the chiral point
$\mu\ell=1$ the left moving generators parameterized by
$\epsilon^-(x^-)$ also become trivial \cite{as}. Hence there is an
enhancement of the trivial symmetry group and the ASG is generated
by a single chiral Virasoro algebra.

\section{Symmetry generators}
In this section we present a refined expression for the symmetry
generators which corrects expressions appearing in some of the
literature.\footnote{The expressions herein were independently found
using a different formalism in \cite{hmt}.} The corrections are
relevant only when the Brown-Henneaux boundary conditions are
violated. This corrected expression is essential for demonstrating
the general equality of the bulk and boundary expressions for the
energy, as well as for the discussion of log gravity in section
\ref{sec:rbc}.

Our expression follows from the covariant formalism
\cite{barnichbrandt,barnichcompere}, which is based on
\cite{Abbott:1981ff}
  and has been developed in great detail for a
wide variety of applications in \cite{barnichstokes}.\footnote{Some
recent discussions of TMG have used the Brown-York
formalism \cite{Brown:1992br}, which was initially developed
for diffeomorphisms which -- unlike those in (\ref{eq:adt}) --
do not have  components normal to the boundary.  For
Brown-Henneaux  boundary conditions
this subtlety turns out to be
irrelevant.  It is, however, relevant when violations of
the Brown-Henneaux boundary conditions are considered. While the Brown-York
formalism could likely be generalized to this case, the
covariant formalism is more highly developed and hence more
convenient.}
 Let $E_{\mu\nu}^{(1)}(h)$ denote  the linearization of the equation of motion
 (\ref{eom}) about AdS$_3$ metric $\bar g$ with respect to a small perturbation $h$ near the boundary. One may then define the one-form
\be\label{kf} \mathcal{K}(\xi,h)\equiv\xi^\mu
E_{\mu\nu}^{(1)}(h)dx^\nu \ee It is shown in  \cite{Deser:2003vh}
that when $\xi$ is a Killing vector\footnote{This formalism was
discussed for general backgrounds in \cite{Bouchareb:2007yx}, and
further generalized in to the case where $\xi$ is not an asymptotic
Killing vector \cite{Compere:2008cv}.  In this case an additional
term appears on the left hand side of (\ref{eomxi}). }
\bea\label{eomxi}\mathcal{K} (\xi,h)=*d*\mathcal{F}(\xi, h).\eea
Here $\mathcal{F}$ is a two form ``superpotential," which is written
out explicitly in \cite{Compere:2008cv}. It was further shown that
the conserved charges associated to the ASG are then given by the
boundary integral  \be Q_{boundary}(\xi)=-{1\over16\pi
G}\oint_{\p\Sigma}*\mathcal{F} .\ee Here $\p\Sigma$ is the boundary
of a spacelike surface $\Sigma$. Integrating by parts gives  the
bulk expression  \be Q_{bulk}(\xi)=-{1\over16\pi G}\int_{\Sigma} *K
= Q_{boundary}(\xi)\ee In this formula $K$ can be taken to be any
smooth extension of the boundary one-form (\ref{kf}) into the
interior. In the coordinates \rref{gads}, we shall see in the next
section that \bea\label{eq:adt}
  Q_{boundary}(\xi)={1\over32\pi\ell G}\oint_{\p\Sigma} d\phi \left[\epsilon^- \left(-2\p_\rho^2h_{--}+4\p_\rho h_{--}
  +2\p_\rho h_{-+}-4h_{-+}+{e^{2\rho}\over4}h_{\rho\rho}\right)\right.\nonumber
  \\+\left.\epsilon^+\left(8h_{++}-8\p_\rho h_{++}+2\p_\rho^2h_{++}+2\p_\rho h_{-+}-4h_{-+}+{e^{2\rho}\over4}h_{\rho\rho}\right) \right]
 \eea
In the above expression (\ref{eq:adt}), we have only assumed that
$h$ falls off fast enough for $Q$ to be finite, but have not used
the Brown-Henneaux boundary condition (\ref{strictbc}).
Asymptotically, the $\rho\rho$ component of the linearized equation
of motion gives \be 2\p_\rho
h_{-+}-4h_{-+}+{e^{2\rho}\over4}h_{\rho\rho}=0\label{cst}\ee
Condition (\ref{cst}) is an asymptotic constraint, as it involves
only the fields and not their time derivatives and hence weakly
vanishes in the Dirac bracket formalism. See \cite{comperedetournay}
for a similar discussion in G\"{o}del spacetime.
 Using the stricter boundary conditions (\ref{strictbc}), and imposing the asymptotic constraints, the expression becomes simply
 \be\label{uuy}
 Q_{boundary}(\xi)={1\over 4\pi\ell G}\oint_{\p\Sigma} d\phi \epsilon^+h_{++}.
 \ee

 These charges can be decomposed into left and right charges
 $Q^L$ and $Q^R$ generating left and right diffeomorphisms $\xi^L(\e^-)$ and $\xi^R(\e^+)$.
 Note that $\e^-$ does not appear in \rref{uuy}, so for Brown-Henneaux boundary conditions the $\xi^L(\e^-)$ are trivial and the left charges vanish.  This implies that the theory is chiral \cite{as}:
 \be
 Q^L\equiv Q(\xi^L)=0.
 \ee
Hence the name chiral gravity.

In the following we will study violations of the asymptotic boundary conditions where the extra terms in
 $Q_{boundary}$ will contribute. In particular, we will encounter situations in which the
 $\p_\rho h_{--}$ term above does not vanish and $Q^L\neq 0$. In this case the left moving charges can be written in a simple gauge invariant form
 \be\label{dt}
Q^L_{boundary}={1\over8\pi G}\oint_{\p\Sigma} \xi^{L\mu}({\mathcal G}^{(1)}_{\mu\nu}-{g_{\mu\nu}\over2}\mathcal{G}^{(1)})dx^\nu,
  \ee
We see that the left charges are nonzero only if the
curvature perturbation does not vanish on the boundary.

\section{Classical perturbation theory}

In this section we will work out the weak field perturbation expansion of
the equations of motion to second order. We start by
expanding the metric around the AdS$_3$ background as \be
g_{\mu\nu}=\bar{g}_{\mu\nu}+h_{\mu\nu}=\bar{g}_{\mu\nu}+h^{(1)}_{\mu\nu}+h^{(2)}_{\mu\nu}+\cdots\ee
The expansion parameter here is the magnitude of the first order fluctuation
$h^{(1)}$.
Inserting this into the full equation of
motion \be\label{eommm} \mathcal{G}_{\mu\nu}+{1\over\mu}C_{\mu\nu}=0 \ee and expanding to first order in the perturbation we see that $h^{(1)}$ must satisfy \be
E^{(1)}_{\mu\nu}(h^{(1)})\equiv\mathcal{G}^{(1)}_{\mu\nu}(h^{(1)})+{1\over\mu}C^{(1)}_{\mu\nu}(h^{(1)})=0\ee
In this and the following equations indices are raised and lowered using the background metric.  The second order perturbation $h^{(2)}$ is found by expanding (\ref{eommm}) to second order
\be\label{seom}
E^{(1)}_{\mu\nu}(h^{(2)})=-E_{\mu\nu}^{(2)}(h^{(1)})\ee
Explicit computation gives the left hand side of (\ref{seom}) \bea\label{eq:e1}
E^{(1)}_{\mu\nu}&=&\mathcal{G}^{(1)}_{\mu\nu}+{1\over2\mu}(\epsilon_\mu\,^{\alpha\beta}\nabla_\alpha\mathcal{G}^{(1)}_{\nu\beta}
+\epsilon_\nu\,^{\alpha\beta}\nabla_\alpha\mathcal{G}^{(1)}_{\mu\beta})\\
\mathcal{G}^{(1)}_{\mu\nu}&=&R^{(1)}_{\mu\nu}+{2\over\ell^2}h_{\mu\nu}-{1\over2}g_{\mu\nu}(R^{(1)}+{2\over\ell^2}h)\eea
where \bea
R^{(1)}_{\mu\nu}&=&\half(-\nabla^2h^{}_{\mu\nu}-\nabla_\mu\nabla_\nu
h+\nabla^\lambda\nabla_\nu h^{}_{\mu
\lambda}+\nabla^\lambda\nabla_\mu  h^{}_{\nu \lambda})\\
\Gamma^{(1)\lambda}_{\mu\nu}&=&{1\over2}\left[\nabla_\mu
h^{\lambda}_\nu+\nabla_\nu h^{\lambda}_\mu-\nabla^\lambda
h^{}_{\mu\nu} \right] .\eea

The right hand side of (\ref{seom}) is
\bea\label{eq:e2}
E_{\mu\nu}^{(2)}&=&\mathcal{G}^{(2)}_{\mu\nu}+{1\over2\mu}[(
\epsilon_\mu\,^{\alpha\beta}\nabla_\alpha\mathcal{G}^{(2)}_{\beta\nu}+
h_{\mu\lambda}\epsilon^{\lambda\alpha\beta}\nabla_\alpha\mathcal{G}^{(1)}_{\beta\nu}
-{h\over2}\epsilon_\mu\,^{\alpha\beta}\nabla_\alpha\mathcal{G}^{(1)}_{\beta\nu}\\
&~~&
-\epsilon_\mu\,^{\alpha\beta}\Gamma^{(1)\lambda}_{\nu\alpha}\mathcal{G}^{(1)}_{\beta\lambda})+(\mu\leftrightarrow\nu)]\non\\
\mathcal{G}^{(2)}_{\mu\nu}&=&R_{\mu\nu}^{(2)}-{g_{\mu\nu}\over2}(R^{(2)}
-h^{\lambda\sigma}R^{(1)}_{\lambda\sigma}+h^{\lambda
\alpha}h^\sigma_\alpha
R_{\lambda\sigma})-{h_{\mu\nu}\over2}(R^{(1)}+{2\over\ell^2}h)\\R_{\mu\nu}^{(2)}&=&\nabla_\lambda\Gamma^{(2)\lambda}_{\mu\nu}-\nabla_\nu\Gamma^{(2)\lambda}_{\mu
\lambda} +\Gamma^{(1)\lambda}_{\lambda
\sigma}\Gamma^{(1)\sigma}_{\mu\nu}-\Gamma^{(1)\lambda}_{\nu
\sigma}\Gamma^{(1)\sigma}_{\mu \lambda}
\\
\Gamma^{(2)\lambda}_{\mu\nu}
&=&-{h^{\lambda \sigma}\over2}(\nabla_\nu h_{\sigma \mu}+\nabla_\mu
h_{\sigma \nu }-\nabla_\sigma h_{\mu\nu}) . \eea

The one-form $\mathcal K$ in (\ref{kf}) may now be
constructed to second order from $E^{(1)}(h^{(2)})$ and shown to be
the divergence of a two-form $\mathcal F(h^{(2)})$. The resulting
boundary expression  for the charges \be\label{bex}
Q_{boundary}(\xi)= -{1\over16\pi G}\oint_{\p\Sigma}*{\mathcal
F(h^{(2)})}\ee yields the expression quoted in
(\ref{eq:adt}). The bulk expression is then obtained by integrating
by parts. When $\xi$ is a background Killing vector it is straightforward to write this bulk charge explicitly\be
\label{beux} Q_{bulk}(\xi)={1\over16\pi G}\int_{\Sigma} *(\xi^\mu
E_{\mu\nu}^{(2)}(h^{(1)})dx^\nu ).\ee For general $\xi$ one can write a similar but somewhat more complicated expression.

At
this point we have not assumed Brown-Henneaux boundary conditions. We note that  it is crucial that the
$\p_\rho h_{--}$ terms in (\ref{eq:adt}) are included; these terms are omitted in some discussions in the literature. Without them the bulk and boundary
expressions \rref{beux} and \rref{bex} would not be equal.

\subsection{Chirality confinement}

We now turn to a discussion of solutions of the linearized equations
and their second-order back reaction.  One may consider a basis of
eigenmodes of $\xi^L_0=\p_-$ and $\xi^R_0=\p_+$, or equivalently
energy and angular momentum. Such eigenmodes were constructed in
\cite{lss}, where it was shown that all the (non-gauge) modes
obeying the boundary conditions (\ref{strictbc}) have vanishing left
charges and are in the $(h_L,h_R)=(0,2)$ highest weight
representation  of $SL(2,\mathbb{R})_L\times SL(2,\mathbb{R})_R$.
These are the right-moving boundary gravitons and can be constructed
from non-trivial $\xi^R_{-2}$ and $\xi^R_{-1}$ diffeomorphisms on
the AdS$_3$ background. There are also weight $(2,0)$ left-moving
excitations, but these can be eliminated by trivial $\xi^L_{-2}$ and
$\xi^L_{-1}$ diffeomorphisms. This is in contrast with the
situation for generic $\mu$ where there are three types of
eigenmodes in highest weight representations:  chiral left and right
boundary gravitons as well as massive gravitons transforming in a
non-chiral highest weight $\half (3+\mu\ell, -1+\mu\ell)$
representation. As $\mu\ell\to 1$, the weight of the massive
graviton approaches $(2,0)$ and  its wave function degenerates with
that of the  left-moving boundary graviton. Consequently it can also
be eliminated by a trivial diffeomorphism. Hence the disappearance of
the massive and left moving representations at the chiral point  is a direct result of the enhancement of the group of trivial
symmetries.

However, there is no guarantee that all solutions of the linearized
equations obeying the boundary conditions (\ref{strictbc}) have an
expansion in terms of $(\xi^L_0,\xi^R_0)$ eigenmodes, or fall into
highest weight representations.  Interestingly, modes without
such an expansion do exist. Examples were explicitly constructed in
\cite{Giribet:2008bw} (building on results of
\cite{Grumiller:2008qz}) and will be denoted $h^{(1)}_{GKP}$.
$h^{(1)}_{GKP}$ cannot be Fourier expanded as eigenmodes of
$\p_\tau$ because it grows linearly in $\tau$. Moreover the GKP
modes are non-chiral: the quadratic bulk expressions for left
and right moving  charges are non-zero \be \label{gkpc} E_L\equiv
Q^L_{bulk}(\xi^L_0, h^{(1)}_{GKP})=-{\ell\over12G },~~~E_R\equiv
Q^R_{bulk}(\xi^R_0, h^{(1)}_{GKP})=-{\ell\over24G}.\ee

On the other hand, we can also compute the charges from the boundary
expression. This involves first solving for the second order
perturbation $h^{(2)}_{GKP}$ and then evaluating the boundary
integral. Since the bulk and boundary expressions are
equal we must have \be E_L= Q^L_{boundary}(\xi^L_0,
h^{(2)}_{GKP})=-{\ell\over12G}, \ee where we have imposed the
condition (\ref{cst}). This cannot be nonzero if $h^{(2)}_{GKP}$
obeys the boundary condition (\ref{strictbc}). We conclude
that $h^{(2)}_{GKP}$ violates the boundary condition, and there is no exact
solution to chiral gravity already at second order with the
prescribed boundary condition. Explicit expressions for
$h^{(1)}_{GKP}$ and $h^{(2)}_{GKP}$ are given below. $h^{(2)}_{GKP}$
grows linearly at infinity so that $\p_\rho h^{(2)}_{GKP}$ gives a
nonvanishing contribution to the boundary expression for the left
charge.

This resolves the apparent contradiction between the the vanishing of
$E_L$ and the existence of  boundary-condition-obeying solutions of the linearized equations with nonzero $E_L$.
The latter are obstructed at second order and are not the linearization of  boundary-condition-obeying solutions of the exact equations.

In the introduction we made an analogy between non-chiral
solutions of linearized chiral gravity and free quark solutions of
linearized QCD: neither are approximate finite-energy solutions of the
exact theory. An alternate, purely classical,  analogy can be found
in Maxwell electromagnetism coupled to a charged scalar in 1+1 dimensions.
At linear order there are scalar field configurations of order
$\epsilon $ with finite charge and finite energy. However these
disappear from the finite energy spectrum at quadratic order: there
is an electric field of order $\epsilon^2$ which carries infinite
energy due to an infrared divergence.  So there are no finite energy
excitations with nonzero charge: charge is confined. Here we
are finding in analogy that non-chiral excitations are confined. In
the Maxwell case, there is only one conserved quantity -- the electric
charge -- which must vanish.  This implies that the linearized solutions must obey a one parameter constraint in order to approximate exact solutions to the theory. In chiral gravity there are infinitely
many conserved charges $Q^L$ which must vanish. This leads to
infinitely many constraints, and the elimination of an entire (left)
chiral sector of the theory.

We have shown that the linearization of all finite excitations of
chiral gravity must be chiral in the sense that the quadratic bulk
expression for $E_L$ (as well as the other left charges) must
vanish. This is irrelevant to the energy eigenmodes which are in any
case chiral, but it eliminates the nonchiral mode $h^{(1)}_{GKP}$
which, from \rref{gkpc}, has $E_L=-{\ell\over12G}$. In principle
there could be additional modes which are chiral but still have
negative energy $E=E_L+E_R=E_R$. This would ruin perturbative
stability. This seems highly unlikely since all linear chiral modes
are associated with asymptotic symmetries, and we know already that
the ASG is generated by exactly one copy of the Virasoro algebra. This Virasoro algebra is
already accounted for by the $(0,2)$ mode.

For the sake of completeness, in sections (\ref{sec:cdww}) and (\ref{sec:gkp}) we will
compute explicitly the second order perturbation
resulting from the various alleged counterexamples to the chiral gravity conjecture.  We will see the infrared divergence described above and
conclude that the linearization of the exact spectrum consists only of the right-moving boundary gravitons.

We note that it is in principle straightforward to find $all$ solutions of the linearized constraint equations in global coordinates, rather than just the energy eigenmodes described above. However, the analogous computation has already been solved in Poincar\'e coordinates \cite{Carlip:2008jk2}.  So we will work primarily in Poincar\'e coordinates.
We will then show in section (4.4) that on global AdS$_3$ all linearized solutions which are non-singular at second order must be chiral and obey the linearized
Einstein equations.

\subsection{The CDWW modes}\label{sec:cdww}

Carlip, Deser, Waldron and Wise (CDWW) have described
all solutions of the linearized
equations of motion which are smooth in Poincar\'e coordinates
\cite{Carlip:2008jk2}.
These include many nonchiral modes. We will first show that
all of these nonchiral modes are singular at second order on the boundary
of the Poincar\'e patch. All modes which are nonsingular at second
order are chiral and obey the linearized Einstein equation.

We use Poincar\'e
coordinates\footnote{The orientation here is
$\epsilon_{t\phi\rho}=\sqrt{-g}$, or equivalently,
$\epsilon_{+-z}=\sqrt{-g}$.  }\be ds^2={-dt^+dt^-+dz^2\over z^2} \ee
and light-front gauge \be h_{--}^{(1)}=h_{+-}^{(1)}=h_{-z}^{(1)}=0.
\ee
Following \cite{Carlip:2008jk2} we may integrate out $\p_-h_{+z}^{(1)}$ and $\p_-^2h_{++}^{(1)}$ in the quadratic action.  The equation of
motion for $h^{(1)}_{zz}$ becomes \bea
\p_+\p_-h_{zz}^{(1)}&=&{1\over4z^2}[z^2\p_z^2+3z\p_z+(-\mu^2+4\mu-3)]h_{zz}^{(1)}\label{eom2}\eea

The general solution of (\ref{eom2}) is a real linear combination of
the modes \bea h^{(1)}_{\omega^+,\omega^-}&=&\sqrt{\omega\over 4\pi
E}{1\over z}e^{-i(\omega^+t^++\omega^-t^-)}J_{|2-\mu|}(2\omega
z)\label{bsl}\\h^{(1)*}_{\omega^+,\omega^-}&=&\sqrt{\omega\over 4\pi
E}{1\over
z}e^{i( \omega^+t^++\omega^-t^- )}J_{|2-\mu|}(2\omega z)
\eea
where
\bea
\omega^2&=&\omega^+\omega^-,\quad
E={\omega^++\omega^-\over2},\quad
k={\omega^+-\omega^-\over2}\non\eea A general solution can be written as a wave packet \be
h^{(1)}_{zz}=\int d\omega dk[ a(\omega^+,
\omega^-)h^{(1)}_{\omega^+,\omega^-}+a^*(\omega^+
,\omega^-)h_{\omega^+, \omega^-}^{(1)*}]\label{wpacket}\ee
The other components of the linear fluctuation are
obtained from (\ref{wpacket}) by \bea
\p_-h_{+z}^{(1)}&=&\half[\p_z+{-\mu+2 \over
z}]h_{zz}^{(1)}\label{eomp}\\\label{eompp}
\p_-^2h^{(1)}_{++}&=&\half[2\p_+\p_--{\mu\over z} \p_z+{\mu^2-3\mu
\over z^2}]h_{zz}^{(1)}\eea

The left moving charges $Q^L$ can now be computed from the bulk
quadratic expression (\ref{eq:e2}).  They are in general nonzero. For example \bea E_L&=&-{1\over128\pi G}\int dz dx\left[ z^3
\left((\p_zh^{(1)}_{zz})^2+4(\p_-h^{(1)}_{zz})^2\right)\right.\\&~&+\half\left.\p_z\left({z^2}(9+z\p_z)(h^{(1)}_{zz})^2\right)\right]\label{el}
\non\\&=&-{1\over128\pi G}\int dz dx\left[ 4z^3
\left(-\p_+\p_-h_{zz}^{(1)}h_{zz}^{(1)}+(\p_-h_{zz}^{(1)})^2]\right)\right.\\&&+ \half\p_z\left.\left(z^2({9}+z\p_z)(h^{(1)}_{zz})^2\right)\right]\non\eea
We have discarded here total derivatives of $t^-$ which vanish upon integration over $x$. This expression
is a total derivative plus a negative semi-definite term. \bea
E_L&=&-{1\over32\pi G}\int dz dx z^3
\left[-\p_+\p_-h_{zz}^{(1)}h_{zz}^{(1)}+(\p_-h^{(1)}_{zz})^{2}\right]\\&=&-{1\over64\pi
G}\int d\omega dk~~ \omega^-~|a(\omega^+,\omega^-)|^2\eea
This vanishes if and only
the mode has support in the region \be w^-=0,\ee In this case $h^{(1)}$ obeys the linearized Einstein equation\be {\mathcal G }^{(1)}_{\mu\nu}(h^{(1)})=0. \ee

In order to make this completely explicit, we will now compute the second order perturbation of the CDWW modes. We will compute the curvature rather than the metric, as divergences in the latter can be coordinate artifacts.
 The $--$ component of (\ref{seom}) is \bea\label{ffv} &&\p_z
{\mathcal G }^{(1)}_{--}(h^{(2)})-\p_-{\mathcal G
}^{(1)}_{z-}(h^{(2)})\\ \non &=&{1\over2} z^3\left[
\p_-h^{(1)}_{zz}\p_-h^{(1)}_{zz}
+\p_-\left(2z\p_-(h^{(1)}_{+z}\p_-h^{(1)}_{zz})-2zh^{(1)}_{zz}\p_-^2h^{(1)}_{+z}-\p_-h^{(1)}_{zz}(z\p_z+{5\over2})h^{(1)}_{zz}\right)\right].
\non\eea A boundary condition for this differential equation is
obtained by noting that at the boundary point $z=\infty$,
\rref{strictbc} implies ${\mathcal G }^{(1)}_{--}(h^{(2)};\infty,
t^+,t^-)=0$. 
If $h^{(1)}$ is one of the $\omega_+,\omega_-$
eigenmodes, equation \rref{ffv} decomposes into three equations which depend
on $t^\pm$ as either $e^{\pm 2i(\omega_+t^+ +\omega_-t^-)}$ or are
constant in $t^\pm$.  Consider the constant piece, for which  ${\mathcal G
}^{(1)}_{z-}(h^{(2)})$ vanishes by  symmetry. We may then solve for
the constant part of  ${\mathcal G }^{(1)}_{--}(h^{(2)})$:
\be\label{opp} {\mathcal G }^{(1)}_{--}(h^{(2)};z,
t^+,t^-)=-\half\omega_-^2\int_{-\infty}^z  dz' z'^3
h_{\omega^+,\omega^-}^{(1)*}h_{\omega^+,\omega^-}^{(1)} \ee which is
strictly negative unless $\omega^-=0$. Thus it is impossible
for ${\mathcal G }^{(1)}_{--}(h^{(2)})$ to vanish everywhere on the
boundary $z=0$ and $z=\infty$ unless $\p_-h^{(1)}$ itself vanishes
everywhere. This leaves only the chiral $\omega^-=0$ modes which
solve the linearized Einstein equation.
We see explicitly that the
linearized modes or, since the right hand side is always negative,  any superposition thereof  must obey the
linearized Einstein equation. Looking at the Fourier modes of ${\mathcal G }^{(1)}_{--}(h^{(2)})$ gives more constraints leading again to $\omega^-=0$.

We note that the above expressions for the curvature at second
order, and hence the conclusion that the boundary conditions are
violated, follows directly from the perturbative expansion of the
equations of motion. Thus although our discussion was motivated by charge conservation, our conclusions ultimately do not rely
on any particular expressions for or properties of the charges.

\subsection{The GKP mode}\label{sec:gkp}

A interesting nonchiral solution of the linearized equations
was constructed by Giribet, Kleban and Porrati (GKP) in \cite{Giribet:2008bw2}.
This mode is not an $(\xi^L_0, \xi^R_0)$ eigenmode but nevertheless obeys
the Brown-Henneaux boundary conditions (\ref{strictbc}).  It may be written as
 \be
h^{(1)GKP}_{\mu\nu}={\mathcal L}^R_{-1}(y(\tau,\rho){\mathcal
L}^L_{-2}\bar{g}_{\mu\nu})+\mathcal {L}_\xi \bar{g}_{\mu\nu}\ee
where ${\mathcal L}^{L,R}_n$ is the Lie derivative with respect to
$\xi^{L,R}_n$ and \bea y(\tau,\rho)&=&-i\tau-\ln(\cosh\rho)\\
\xi&=&-{iy(\tau,\rho)\sinh(\rho)\over2\ell^2\cosh^5(\rho)}e^{-i(\tau^++2\tau^-)}\xi^R_0.\eea
The conserved charges are \bea E^L_{GKP}&=&-{\ell\over 12G}\\
E^R_{GKP}&=&-{\ell\over24G}\eea We may now solve asymptotically for
$h^{(2)GKP}$ using (\ref{seom}), which reduces to \bea
E^L&=&{1\over16\pi\ell G}\oint d\phi (2\p_\rho
h^{(2)}_{--}-\p_\rho^2
h^{(2)}_{--})=E^L_{GKP}\label{eq:el}\\E^R&=&{1\over 16\pi\ell
G}\oint d\phi (4h^{(2)}_{++}-4\p_\rho h^{(2)}_{++}+\p_\rho^2
h^{(2)}_{++})=E^R_{GKP}  \label{eq:er}\eea
The $\phi$ independent
solution is \bea h^{(2)}_{--}&=&
 4G\ell E^L \rho+\cdots\label{eq:hpp}
\\h^{(2)}_{++}&=&2G\ell E^R+\cdots
\label{eq:hmm}\eea  where $\dots$ denotes terms which are subleading in $\rho$. In particular, for the GKP modes, we have \bea
h^{(2)GKP}_{--}&=& -{\ell^2\rho\over3}+\cdots\label{eq:hpp}
\\h^{(2)GKP}_{++}&=&-{\ell^2\over12 }+\cdots
\label{eq:hmm}\eea
From (\ref{eq:hpp}) we see that the Brown-Henneaux boundary conditions  (\ref{strictbc}) are
violated. We conclude that $h^{(1)GKP}$ is not the
linearization of an exact solution to the equations of motion.

\subsection{Global modes}
We can now argue that {\it all} solutions to the linearized equations of motion that obey Brown-Henneaux boundary conditions at second order must be solutions of the linearized Einstein equations.  In particular, we are left only with the right-moving boundary gravitons.

To prove this, one could study the linearized equations of motion in global rather than Poincar\'e coordinates.  However this can be avoided by noting that every mode which is smooth and
asymptotically AdS$_3$ in global coordinates is smooth on the
Poincare patch and hence has an expansion in CDWW
modes of section (\ref{sec:cdww}).\footnote{Of course, the converse is not true:  modes which are well
behaved on the Poincare patch may not be well-behaved globally.} We
have seen that of these modes only the ones with vanishing Einstein
tensor obey Brown-Henneaux boundary conditions at second order.  As the right hand side of
\rref{opp} is negative definite, we cannot cancel this divergence for any
linear superpositions of modes. Hence all global modes must obey the
linearized Einstein equation.
\section{A Birkhoff-like theorem}

We have seen that any solution of chiral gravity is, at the
linearized level, locally AdS$_3$. This might lead one to suspect
that all solutions of chiral gravity are locally AdS$_3$ at the full
non-linear level. In this section we will see that this is indeed
the case for a particularly simple class of solutions: those which
are stationary and axially symmetric.  For this class of solutions
the full non-linear equations of motion, although still surprisingly
complicated, are reasonably tractable. We will conclude that, once
we impose Brown-Henneaux boundary conditions, the only solutions are
the BTZ black holes.

A similar result was obtained for general values of $\mu$ by
\cite{Aliev:1996eh, Cavaglia:1999si}, who made the somewhat stronger assumption of a hypersurface orthogonal
Killing vector field.

\subsection{Stationarity and axial symmetry}

We start by studying the equations of motion of TMG for stationary,
axially symmetric spacetimes, following the approach of
\cite{Moussa:2003fc}.

A three dimensional spacetime with two commuting $U(1)$ isometries may, through judicious choice of coordinates, be written in the form
 \begin{equation} \label{ansatz}{ds^2}=-X^+(\sigma)d\tau^2+X^-(\sigma)d\phi^2+2Y(\sigma)dt d\phi+{d\sigma^2\over
 X^+(\sigma) X^-(\sigma)+Y(\sigma)^2}\end{equation}
The two $U(1)$ isometries are the generated by Killing vectors $\p_\tau$ and $\p_\phi$.
We are interested in axially symmetric solutions, so we will take the angular direction to be periodic $\phi\sim\phi+2\pi$.
We have chosen the coefficient of $d\sigma^2$ for future convenience.

The geometry of the solution is encoded in the three functions $X^\pm(\sigma), Y(\sigma)$, which we will package into a three dimensional vector $ {\bf X}$ with components $X^i$, $i=0,1,2$ given by
\begin{equation}
X^0 = {X^+(\sigma) - X^-(\sigma)\over 2},~~~~~
X^1 = {X^+(\sigma) + X^-(\sigma)\over 2},~~~~~X^2=Y
\end{equation}
The dynamics of stationary, axial metrics in three dimensions may be thought of as the dynamics of a particle with position ${\bf X}(\sigma)$ moving in the auxiliary space $\R^{2,1}$ parameterized by $\bx$.

For the ansatz (\ref{ansatz}), the equations of motion of TMG are
\begin{equation}
-2\mu\textbf{X}''=2\textbf{X}\times\textbf{X}'''+3\textbf{X}'\times\textbf{X}''\label{eomx}
\end{equation}
\begin{equation}
4 = \textbf{X}'^2 -{2\over \mu} \textbf{X}\cdot(\textbf{X}'\times\textbf{X}'')\label{constraintx}
\end{equation}
We have set $\ell=1$ for convenience.
Here $'$ denotes $\p_\sigma$ and we have defined the Lorentz invariant dot product and cross product\footnote{In Lorentzian signature some of the usual cross product identities must be altered (e.g. ${\bf A}\times ({\bf B} \times {\bf C})  = {\bf C} ({\bf A} \cdot {\bf B}) - {\bf B} ({\bf A} \cdot {\bf C})$ differs by a sign from the usual case).}
\be \textbf{A}\cdot
\textbf{B}\equiv\eta_{ij}A^iB^j,~~~~\hbox{and}~~~~~
(\textbf{A}\times \textbf{B})^i\equiv
\eta^{il}\epsilon_{ljk}A^jB^k,\quad \epsilon_{012}=1\ee

In order to understand these equations, it is helpful to note that for our ansatz (\ref{ansatz}) the action of TMG is
\begin{equation} I={1\over16\pi G}\int d\sigma
{1\over2}\left(\bx'^2-{1\over\mu}\bx\cdot(\bx'\times
\bx'')\right)\label{action}\end{equation}
This is the action of a Lorentz invariant particle mechanics in $\R^{2,1}$.
Equation (\ref{eomx}) is found by varying this reduced action with respect to $\bf X$.  Equation (\ref{constraintx}) represents an additional constraint which arises due to gauge-fixing.  In fact, the right hand side of (\ref{constraintx}) is just the conserved Hamiltonian of the reduced action (\ref{action}).

The equation of motion (\ref{eomx}) can easily be integrated once.
To see this, note that tha action \rref{action} is invariant under Lorentz transformations in $\R^{2,1}$.  Hence there is a conserved angular momentum ${\bf J}$ associated to these Lorentz transformations, which we can compute using the Noether procedure (taking into account the higher derivative terms)
\begin{equation}\textbf{J}={1\over 16 \pi G} \left(\bx\times\bx'
 -{1\over2\mu}[\bx'\times(\bx\times\bx')-2\bx\times(\bx\times\bx'')]\right)\label{angularx} \end{equation}
One can check explicitly that (\ref{angularx}) is the the first
integral of (\ref{eomx}). The dynamics of the system is given by the set of second order differential equations (\ref{angularx}) and (\ref{constraintx}).  The auxiliary angular momentum $\bf J$ should not be confused with the physical angular momentum of the spacetime, although we shall see that they are closely related.
With the help of (\ref{eomx}) and (\ref{constraintx}) we can write (\ref{angularx}) as a second order differential equation for ${\bf X}''$:
\begin{equation}
2 \bx^2{\bf X}'' = {32\pi G\mu {\bf J} - 2\mu {\bf X} \times {\bf X}' + {\bf X}' ({\bf X} \cdot {\bf X}')+ {\bf X} (6-{5\over 2} {\bf X}'{}^2)}
\label{eomfinal}
\end{equation}
This equation can then be integrated to give a solution $\bx(\sigma)$ to the equations of motion.  For future reference, we note that with the help of the equations of motion we can write the Hamiltonian constraint as
\be
(\bx'^2 -4)={2\over \mu} \bx\cdot (\bx'\times\bx'') = -{4\over 3} \bx \cdot \bx''
\ee

\subsection{Boundary conditions}

We will now consider axially symmetric, stationary solutions of TMG obeying Brown-Henneaux boundary conditions.

We start by noting that global metric on AdS$_3$ can be written as
\begin{equation}
ds^2=-(2\sigma+1)d\tau^2+2\sigma d\phi^2+{d\sigma^2\over 2\sigma(2\sigma+1)}
\label{ads}
\end{equation}
which is of the form (\ref{ansatz}) with \begin{equation}
{\bf X}_{AdS} = 2\sigma(0,1,0)+ {1\over 2}(1,1, 0)
\end{equation}
Here $\sigma$ is related to the usual global radial coordinate on AdS$_3$ by  $\sigma = {1\over 2} \sinh^2 \rho$.
The asymptotic boundary is at $\sigma\to \infty$.
Likewise, the BTZ black hole can be written in the form (\ref{ansatz}) with
\begin{equation}
\bx_{BTZ}  = 2\sigma(0,1,0) - 4 G M (1,1,0) + 4 G J(0,0,1)
\label{btz}
\end{equation}
Here $M$ and $J$ are the ADM mass and angular momentum of the BTZ black hole in Einstein gravity.
As we are working in units with $\ell=1$, empty AdS$_3$ has energy $ M=-1/8G$.

Let us now consider an arbitrary metric obeying Brown-Henneaux boundary conditions.  Comparing with the AdS metric (\ref{ads}) one can check that a metric of the form (\ref{ansatz}) obeys Brown-Henneaux boundary conditions if
\begin{equation}
{\bf X} =  2\sigma (0,1,0) + {\cal O}(1) ,~~~~~~{\rm as}~~\sigma\to\infty
\end{equation}
In analogy with (\ref{btz}) we will write this boundary condition as
\begin{equation}
{\bf X} = 2(\sigma-\sigma_0)(0,1,0) - 4 G M (1,1,0) + 4 G J(0,0,1) +
\dots \label{bhbc}
\end{equation}
where $\sigma_0$, $M$ and $J$ are constants, $G$ is Newton's constant and $\dots$ denotes terms which vanish as $\sigma\to\infty$.
By comparing with (\ref{btz}), we see that $M$ and $J$ are the usual ADM mass and angular momentum of the spacetime as measured at asymptotic infinity in Einstein gravity.  The parameter $\sigma_0$ is just a shift in the radial coordinate and does not have a coordinate-independent meaning.

Let us now consider solutions to the equations of motion of TMG with the boundary conditions (\ref{bhbc}).  As the angular momentum $\bf J$ is a constant of motion we can compute it at $\sigma\to \infty$.  Plugging (\ref{bhbc}) into (\ref{angularx}) we find
\begin{equation}
2 \pi ~ {\bf J} = (J,0,-M)+{1\over\mu} (-M,0,J)
\end{equation}
The auxiliary angular momentum $\bf J$ is just a rewriting of the usual mass and angular momentum of the solution.  We note that
\begin{equation}
(2\pi {\bf J})^2 = \left(1-{1\over \mu^2}\right)\left(M^2 - J^2\right)
\end{equation}
For values of $\mu>1$, we see that the angular momentum vector is spacelike for solutions obeying the cosmic censorship bound $M>J$.  For extremally rotating solutions $\bf J$ is null, ${\bf J}^2=0$.

Finally, we turn to the case of chiral gravity ($\mu=1$) with Brown-Henneaux boundary conditions.  In this case the angular momentum vector is always null:
\begin{equation}
2\pi {\bf J} = (J-M)(1,0,1),~~~~~{\bf J}^2 = 0
\end{equation}
This property will turn out to be very useful.

\subsection{Solutions }

We will now specialize to chiral gravity ($\mu=1$) and study axially symmetric, stationary solutions obeying Brown-Henneaux boundary conditions.  We will demonstrate that if the spacetime has a single asymptotic boundary obeying Brown-Henneaux boundary conditions, then the spacetime must be locally AdS$_3$. We will also assume that $\bx (\rho)$ is an analytic function of $\rho$.

At $\sigma\to \infty$, the vector $\bx$ is spacelike.  As long as $\bx$ remains spacelike, we can continue to smoothly evolve our metric into the interior using (\ref{eomfinal}).   In fact,  $\bx$ must become null  -- with $\bx^2=0$ -- for some finite value of $\sigma$.  To see this, consider what would happen if $\bx^2$ remained strictly positive for all values of $\sigma$.  In this case the evolution equation (\ref{eomfinal}) would allow us to evolve $\bx$ all the way to $\sigma\to-\infty$.  The region $\sigma\to-\infty$ then represents an additional asymptotic boundary.  To prove this, note that in order for $\bf J$ to remain finite at $\sigma\to -\infty$, $\bx^2$ must either remain finite or diverge no more quickly than $\sigma^2$.  The line element
\begin{equation} ds^2 \sim {d\sigma^2 \over \bx^2} +\dots \end{equation}
then implies that points with $\sigma\to-\infty$ lie an infinite proper distance from points with finite $\sigma$.  As we are assuming Brown-Henneaux boundary conditions with a {\it single} asymptotic boundary, we must not allow this additional boundary at $\sigma\to-\infty$. We conclude that there must be finite value of $\sigma$ where $\bx$ becomes null, i.e. $\bx^2=0$.

We will now proceed to study the equation of motion near the point where $\bx^2=0$.  First, let us shift our coordinate $\sigma$ so that this point occurs at $\sigma=0$.  We will assume that the metric is analytic at this point, so is equal to its Taylor expansion
\begin{equation}
{\bf X} = \sum_{n \ge 0} {1\over n!} \sigma^n {\bf X}_n
\label{taylor}
\end{equation}
The coefficients $\bx_n$ are finite and given by derivatives of $\bf X$ at $\sigma=0$, with $\bx_0^2=0$.  In fact, one can check that the point $\sigma=0$ is either a horizon or an origin of polar coordinates, depending on the relative values of $\bx_0$ and $\bx_1$.

We now turn to the equations of motion.
By plugging (\ref{taylor}) into (\ref{eomfinal}) and expanding order by order in powers of $\sigma$ we obtain a set of  recursion relations which determine the $\bx_n$ in terms of $\bx_0$ and $\bx_1$.  We will now proceed to show that these recursion relations imply that all the terms with $n\ge 2$ in the Taylor expansion (\ref{taylor}) vanish.  This will imply that our Taylor expansion converges, hence the solution can be smoothly matched on to the metric at infinity.  Indeed, comparing with equation (\ref{btz}) we see that our solution
\begin{equation}
\bx = \bx_0 + \sigma \bx_1
\label{fin}
\end{equation}
is simply the BTZ black hole.\footnote{In comparing (\ref{fin}) with (\ref{btz}) we must remember that in (\ref{fin}) we have shifted $\sigma$ to put the horizon at $\sigma=0$, in contrast to (\ref{btz}).}  This allows us to conclude that our solution is locally AdS$_3$.

In order to demonstrate this, let us now expand equation (\ref{eomfinal}) order by order in $\sigma$.
The order $\sigma^0$ term just fixes the angular momentum vector
\begin{equation}
32 \pi G {\bf J} = 2 \bx_0 \times \bx_1 - \bx_1 (\bx_0 \cdot \bx_1)-\bx_0(6-{5\over 2} \bx_1^2)
\end{equation}
In chiral gravity, $\bf J$ must be null, so that
\begin{equation}
(32 \pi G {\bf J})^2 = -4 (\bx_0 \cdot \bx_1)^2(\bx_1^2-4) =0
\label{Jeq}
\end{equation}
implying that either $\bx_1^2-4$ or $\bx_0 \cdot \bx_1$ vanish. We will consider the following cases separately:

\

\noindent
{\it Case 0: $\bx_0=0$}

\

In this case ${\bf J}=0$ and it is easy to prove directly that all the higher order terms vanish.  In particular, (\ref{angularx}) implies that
\be
{\bf J} \cdot \bx = \bx^2 \bx'{}^2 - (\bx\cdot\bx')^2=0
\ee
so that $\bx \times \bx'$ is null.  We also see that
\be
{\bf J}\cdot (\bx \times \bx'-2\bx) = \bx^2 (\bx'^2 -4) = 0
\ee
so that $\bx'^2 =4$.  Thus in the region where $\bx^2$ is positive, $\bx$ and $\bx'$ are spacelike vectors whose cross product $\bx \times \bx'$ is null.
One can use this condition to show that $\bx$ and $\bx'$ obey\footnote{For any two spacelike vectors ${\bf A}$ and ${\bf B}$ whose cross product is null one has the identity ${\bf A} \times {\bf B} = \pm \sqrt{{\bf A}^2} {\bf B} \pm \sqrt{{\bf B}^2}{\bf A}$ where the signs depend on the relative orientations of the vectors.  We have fixed the signs here by comparing to the behavior at asymptotic infinity (where $\bx$ approaches that of an extremally rotating BTZ black hole with $M=J$).}
\be
\bx \times \bx' = -\sqrt{\bx'^2} \bx + \sqrt{ \bx^2} \bx'
\ee
Plugging these identities into (\ref{eomfinal}) we conclude that $\bx''=0$.  Thus all of the higher order terms in the Taylor expansion vanish and the solution is just the BTZ black hole.

\

\noindent
{\it Case 1: $\bx_1^2=4$ and $\bx_0\cdot \bx_1\ne0$ }

\

In this case we must work a little harder and examine the terms in the Taylor expansion (\ref{taylor}) term by term.
The equation for $\bx_2$ is found by expanding (\ref{eomfinal}) to linear order in $\sigma$:
\begin{equation}
3 \bx_0\cdot \bx_1 \bx_2 + 2 \bx_0 \times\bx_2-\bx_1 \bx_0\cdot\bx_2 +5 \bx_0 \bx_1\cdot\bx_2 =0
\label{x2is}
\end{equation}
The Hamiltonian equation expanded to order $\sigma^0$ is
\begin{equation}
\bx_0 \cdot \bx_2=0
\end{equation}
In fact, $\bx_2=0$ is the only solution to this equation.  To see this, note that since $\bx_0\cdot \bx_1=0$ the vectors $\bx_0$, $\bx_1$ and $\bx_0\times \bx_1$ form a basis for $\R^{2,1}$.  So we may expand
\be
\bx_2 = a \bx_0 + b \bx_1 + c \bx_0 \times \bx_1
\ee
for some constants $a,b,c$.  Plugging into the equations of motion we find that each of these constants must vanish, so $\bx_2=0$.

We will now prove by induction that all of the higher order terms in the expansion (\ref{taylor}) must vanish as well.  Let us start by assuming that all of the quadratic and higher terms in the expansion (\ref{taylor}) vanish up to a given order $m$.  That is, let us assume that
\begin{equation}
\bx = \bx_0 + \sigma \bx_1 + \sum_{n\ge m}{1\over n!} \sigma^n \bx_n
\label{hypo}
\end{equation}
for some $m\ge 2$.  Expanding equation (\ref{eomfinal}) to order $\sigma^{m-1}$ gives
\begin{equation}
(4m-5) \bx_0 \cdot \bx_1 \bx_m +2 \bx_0\times \bx_m-\bx_0\cdot \bx_m \bx_1+ 5 \bx_1\cdot \bx_m \bx_0 =0
\label{xmis}
\end{equation}
and the Hamiltonian constraint at order $\sigma^{m-2}$ gives
\be
\bx_0 \cdot \bx_m=0
\ee
Expanding $\bx_m$ in the basis as above, one can again show that each term in the basis expansion vanishes separately. Hence $\bx_m$ vanishes and the inductive hypothesis (\ref{hypo}) holds up to order $m+1$.  In the previous paragraph we proved the case $m=2$, so by induction it follows that that {\it all} $\bx_n$, $n\ge2$ must vanish.

\

\noindent
{\it Case 2: $\bx_0\ne 0$ but $\bx_0 \cdot \bx_1=0$}

\

This special case is a bit more complicated.  We note that since $\bx_0$ is null and $\bx_0\cdot \bx_1=0$ it follows that
\be
\bx_0 \times \bx_1 = \pm \sqrt{\bx_1^2} \bx_0
\label{nullident}
\ee
Expanding the Hamiltonian constraint to order $\sigma^0$ gives
\be
\bx_1^2 - 4 = 2 \bx_2 \cdot (\bx_0 \times \bx_1) = \pm 2 \sqrt{\bx_1^2} \bx_2 \cdot \bx_0
\ee
Comparing to the second form of the Hamiltonian constraint
\be
\bx_1^2-4= -{4\over 3} \bx_2\cdot \bx_0
\ee
we see that either $\bx_1^2 = 4/9 $ or $\bx_1^2=4$.  We will consider these two cases separately.

If $\bx_1^2=4$ then we can show that all higher order terms in the Taylor expansion vanish.  The Hamiltonian constraint is $\bx_2\cdot \bx_0=0$.   Expanding the equation of motion to linear order in $\sigma$ gives equation (\ref{x2is}), which in this case has the solution
\begin{equation}
\bx_2 = a \bx_0
\end{equation}
for some constant $a$.  However, we find that the equation at order $\sigma^3$ implies that $a=0$ so that $\bx_2=0$.  Likewise, equation (\ref{xmis}) as a solution of the form
\begin{equation}
\bx_m = a_m \bx_0
\end{equation}
but the coefficient $a_m$ is set to zero by the equation of motion at order $\sigma^{m+1}$.  Proceeding in this manner we conclude that all of the $\bx_n$ must vanish for $n\ge2$, so the solution is just the BTZ black hole.

Finally, let us consider the case where $\bx_1^2 = 4/9$.  In this case the Hamiltonian constraint
\be
\bx_2 \cdot \bx_0 = {8\over 3}
\ee
implies that $\bx_2$ is non-zero.  Aside from this small difference, the argument proceeds exactly as above.  Expanding the equations of motion order by order in $\sigma$ we discover that all of the terms in the Taylor expansion vanish except for $\bx_0$, $\bx_1$ and $\bx_2$, which obey
\be
\bx_0^2 = \bx_2^2 = \bx_0\cdot \bx_1 =\bx_1\cdot\bx_2= 0,~~~\bx_1^2 = 4,~~~\bx_0\cdot\bx_2 = {8\over 3},~~~\bx_0\times\bx_2=\pm 4 \bx_1.
\ee
One may check that the resulting solution
\be
\bx = \bx_0 + \bx_1 \rho + {1\over 2} \bx_2 \rho^2
\ee
is warped AdS$_3$.  This solution does not obey Brown-Henneaux boundary conditions, because of the ${\cal O}(\rho^2)$ behavior at asymptotic infinity.

This completes the proof.

\section{Log gravity}\label{sec:rbc}

In this section we consider log gravity, which differs from chiral
gravity in that the boundary conditions are relaxed to allow certain
types of growth linear in $\rho$ (and logarithmic in the proper
radius) at infinity. The solutions of log gravity will have energies
which are unbounded below as well as unbounded above. Nevertheless,
the theory is of considerable interest as it contains a novel and
mathematically natural class of solutions  \cite{AyonBeato:2004fq,
cdww, Gibbons:2008vi, Garbarz:2008qn} excluded from chiral gravity.
Here we will show that log gravity is consistent insofar as the
expressions for the conserved charges are finite.  However, the left
charges are in general nonzero, so log gravity is not chiral.  This
result is in agreement with \cite{hmt}.\footnote{However, this
result is not in agreement with reference \cite{Grumiller:2008es},
which neglected a term in the generators.} Moreover, we shall see
that log gravity contains within it a decoupled superselection
sector which is identical to chiral gravity. The relation between
this chiral $Q^L=0$ subsector and the full spectrum of log gravity
is reminiscent of the  relation between the physical states of a
first-quantized string (or any 2d gravity theory) and the larger
Hilbert space including longitudinal modes and ghosts. Indeed,
logarithmic CFTs appeared in the 2D gravity context in \cite{bilal}.

\subsection{Boundary conditions and non-chirality}

The starting point for the development of log gravity was the observation by Grumiller and Johansson (GJ) in that a solution to TMG
at the chiral point can be obtained as \cite{Grumiller:2008qz}  \be\label{hbb} h^{GJ}\equiv
\lim_{\mu\ell \to 1} \frac{h^\mu-h^L}{\mu\ell-1}\,.\ee Here $h^\mu$
and $h^L$ are the wave functions for the massive and left-moving gravitons, respectively.
The mode $h^{GJ}$ is a solution of the linearized
equations of motion, but it is not an energy eigenstate and was not considered in \cite{lss}.  Moreover $h^{GJ}$  does
not obey the Brown-Henneaux boundary conditions as certain
components grow linearly in $\rho$ at the boundary.

 GJ then proposed that the Brown-Henneaux boundary conditions (\ref{strictbc}) be relaxed to allow metric fluctuations to grow linearly as $\rho\to\infty$ \cite{Grumiller:2008qz}. The mode $h^{GJ}$ would be included in the spectrum of such a theory.  However, this proposal does not lead to a consistent theory, because for the general such asymptotic perturbation the right-moving charge $Q^R$
is linearly divergent and hence ill-defined. A modified approach
\cite{as} is to impose chiral boundary conditions for which $h_{--}$
is allowed to grow linearly in $\rho$ but $h_{++}$ or $h_{+-}$ are
not.  Specifically, we take \cite{hmt, Grumiller:2008es}
\be\label{eq:relaxbc} \left(
  \begin{array}{ccccc}
 h_{++}= \mathcal{O}({ 1}) & h_{+-}= \mathcal{O}({1})  &h_{+\rho}= \mathcal{O}(e^{-2\rho})  \\
 & h_{--}=\mathcal{O}(\rho)&h_{-\rho}= \mathcal{O}(\rho e^{-2\rho} )  \\  &  & h_{\rho\rho}= \mathcal{O}(e^{-2\rho}) \\
  \end{array}
\right) \ee The allowed diffeomorphisms are \bea \xi^+&=&\epsilon^+(x^+)+{2e^{-2\rho}}\p_-^2\epsilon^-(x^-)+\cdots \\
\xi^-&=&\epsilon^-(x^-)+{2e^{-2\rho}}\p_+^2\epsilon^+(x^+)+\cdots \\
\xi^\rho&=&-{1\over2}\left(\p_+\epsilon^+(x^+)+\p_-\epsilon^-(x^-)\right)+\cdots
\eea
The leading terms give two sets of Virasoro generators. The subleading terms are trivial and do not appear in the charges.  It is straightforward to see that both $Q^L$ and $Q^R$ are finite for all elements of the ASG. However, since $\p_\rho h_{--}
\neq 0$ we find that
 \be Q^L\neq 0,
 \ee
 so that log gravity is $not$ chiral.

 This opens up the possibility \cite{Grumiller:2008qz}
 that log gravity could be holographically dual to a
logarithmic CFT.  In fact, a logarithmic CFT can never be chiral.\footnote{We thank M. Gaberdiel for pointing this out.}  Moreover it is not hard
to see that the mode \rref{hbb} lies in an indecomposable Virasoro
representation
 (containing the left-moving highest weight representation of massless gravitons) characteristic of a logarithmic CFT \cite{log}.

 While we have seen that the charges are finite for log gravity, more work must be done to show that they actually generate the asymptotic symmetry group, or indeed if log gravity has a canonical formulation at all. A canonical formulation requires the construction of  a closed invertible symplectic form $\Omega$, or equivalently a Dirac bracket, on the physical phase space, The Dirac bracket is  nonlocal in space and its construction involves inverting the constraints. Whether or not the constraints
can be inverted depends on the boundary conditions, and so there is in general no guarantee that Dirac brackets exist for any boundary condition.  Since there are physical zero norm states in log gravity,
invertibility is not manifest.  An elegant covariant construction of $\Omega$ was given for general relativity in  \cite{cw} in the form of an integral $\Omega= \int d\Sigma_\alpha J^\alpha $ over a spatial slice, with $ J^\alpha=\delta \Gamma^{\alpha}_{\nu \lambda} \wedge [\delta g^{\nu \lambda}
+\half g^{\nu \lambda}\delta \ln g]-\delta \Gamma^\lambda_{\nu \lambda }[\delta g^{\alpha \nu}
+\half g^{\alpha \nu}\delta \ln g]$ and $\delta$ the exterior derivative on the phase space. For TMG, there is an additional term proportional to ${1\over \mu}\e^{\alpha \lambda \nu}
\delta \Gamma^\sigma_{\lambda \rho}\wedge \delta \Gamma^\rho_{\nu\sigma}$. It would be interesting to see by direct computation if this symplectic form is both finite and invertible for log gravity.
\subsection{Decoupling the chiral gravity superselection sector }

Log gravity in and of itself does not seem of so much interest because it is not unitary.  Nonunitary theories of quantum gravity are generally easy to construct, and are not expected to shed much led on the presumably unitary theory which describes our four dimensional world.  What makes log gravity interesting is that it contains chiral gravity, which has the possibility of being unitary, within it. The structure of this embedding is intriguing and could be useful for a full understanding of chiral gravity.  In this section we explain how this embedding works.

Let $Q^L_n$ denote the left Virasoro charges.  The classical computation
of the central charge is insensitive to the boundary conditions as
long as the charges are well defined. Therefore the Dirac bracket
algebra \be \label{cvb} \{Q^L_m, Q^L_n\} =i(m-n)Q^L_{m+n},\ee has \be
c_L=0 \ee
as is
the case for chiral gravity. The charges $Q^L_n$ are  conserved for all $n$.
Therefore we can consistently truncate to the
charge superselection sector of the theory with $Q^L_n=0$.
$Q^L_n$ are the Fourier transforms of the linearly growing terms in
$h$
\be
\p_\rho h_{--}=4 \ell G\sum_nQ^L_ne^{in\phi}.
\ee
Therefore in the $Q^L_n=0$ superselection sector  we have \be
\p_\rho h_{--}=0.
\ee
This condition reduces the log gravity boundary conditions \rref{eq:relaxbc} to the chiral gravity boundary conditions \rref{strictbc}.\footnote{ If present, terms in $h_{-\rho}$ proportional to $\rho e^{-2\rho}$ may then, as in \cite{Giribet:2008bw2} be eliminated by a trivial diffeomorphism along $\xi=-2e^{-2\rho}h_{-\rho}\p_+$.} Therefore the $Q^L_n=0$ superselection sector of log gravity is precisely chiral gravity. Charge conservation guarantees that time evolution
preserves the chiral boundary conditions and chiral gravity completely decouples from log gravity. Note that this result is nonperturbative.

At the classical level, this shows that solutions of chiral gravity cannot smoothly evolve into geometries with logarithmic behavior at infinity. Of course, we have not proven cosmic censorship so we cannot rule out
singularities on the boundary for either log or chiral gravity.

One may phrase the issue of classical decoupling of chiral gravity
in a different way in perturbation theory, where one can see the
decoupling by direct computation without invoking charge
conservation.  If we excite two linearized modes of chiral gravity,
will a log mode be excited at the next order? Do the chiral modes
source the log modes? This question has already been answered to
second order in our perturbative analysis of section 4. It is
immediate from inspection of \rref{ffv}-\rref{opp} that if we take
$h^{(1)}$ to solve the linearized Einstein equation, then at second
order  ${\cal G}^{(2)}$ vanishes. Of course one can always add a log
mode obeying the homogenous equation at second order, but as this is
not required the log mode can be decoupled.

This analysis can be extended to all orders. If $h^{(1)}$ is nontrivial and is a linearized solution of chiral gravity, it also solves the linearized Einstein equation,
and is an infinitesimal  nontrivial diffeomorphism. The exact all-orders corrected solution is then just the  finite diffeomorphism.  This obviously is a solution of chiral gravity with no log modes excited.

At the quantum level, the question is trickier. Of course we do not know whether or not either theory exists quantum mechanically.  If log gravity does exists as a logarithmic conformal field theory we know it contains chiral gravity as a superselection sector. In perturbation theory, chiral gravity has only the massless gravitons which are dual to the quantum stress tensor current algebra. The OPEs of these fields obviously close and hence decouple from log gravity.  But we do not know if that superselection sector contains more than just the current algebra, or if it is local or modular invariant. Equivalently, we do not know if the black hole microstates are $exactly$ chiral at the quantum level, or contain  small nonchiral charges which are lost in the semiclassical limit. As we shall now see, this is more or less equivalent to the question of whether or not there are extremal CFTs with large central charge. More discussion of this point can be found in the concluding section.

\section{Quantum partition function}

At this point we have seen that with Brown-Henneaux boundary conditions the linearized spectrum of perturbations of chiral gravity around an AdS$_3$ background includes only right moving boundary gravitons.
We will now use this observation to compute the partition function of the quantum theory assuming applicability of the standard Euclidean methodology. Quite nontrivially, we will find that the resulting  partition function has a consistent quantum mechanical interpretation. This can be regarded as evidence that the quantum theory actually exists.

We wish to compute the torus partition function, which is defined as the
generating function \be\label{zdef} Z(\tau) = Tr~ q^{L_0} q^{\bar
L_0},~~~~~~q=e^{2\pi i \tau} \ee encoding the spectrum of the
theory. $Z(\tau)$ may be thought of as a canonical ensemble
partition function at fixed temperature $\beta^{-1}$ and angular
potential $i\theta$ given by $\tau = {1\over 2\pi} (\theta +
i\beta)$.   We will not attempt to perform the trace in \rref{zdef}
directly; this would require an understanding of the Hilbert space
of chiral gravity at the non-perturbative level.  We instead adopt
an indirect approach, following \cite{Maloney:2007ud,Dijkgraaf:2000fq,
 Manschot:2007zb, Yin:2007gv}.

The standard strategy for computing canonical ensemble partitions in quantum field theory is to go to Euclidean signature.  In this case $Z(\tau)$,  originally formulated in terms of the Hilbert space of the Lorentzian theory, takes the form of a path integral
\be\label{Zpath}
Z(\tau) \sim \int Dg e^{-k I[g]}
\ee
The dependence of this path integral on $\tau$ enters through the boundary conditions imposed on the metric $g$; the boundary is taken to be a two torus $T^2$ with conformal structure parameter $\tau$.  For clarity, we have extracted from equation \rref{Zpath} an explicit factor of the dimensionless coupling constant of the theory, $k={\ell/16G}$.  In terms of the central charge of the dual boundary theory, $k=c/24$.

At large $k$ the dominant contribution to the path integral is given by the saddle point approximation
\be\label{Zsum}
Z(\tau) = \sum_{g_{c}} e^{-k I[g_{c}] + I^{(1)}[g_c] + {1\over k} I^{(2)}[g_c] + \dots}
\ee
Here the sum is over classical solutions $g_{c}$ to the equations of motion of the theory.  $I[g_{c}] $ denotes the corresponding classical action.  The subleading terms of the form  $ k^{1-n} I^{(n)}[g_c]$ represent quantum corrections to the effective action at $n^{th}$ order in perturbation theory.

We will take equation \rref{Zsum} to be our working definition of
the path integral of chiral gravity \rref{Zpath} and assume the
equivalence of \rref{Zpath} with \rref{zdef}.  In quantum mechanics,
this equivalence  can be rigorously established. In quantum field
theory in general it cannot be proven, but has worked well in many
situations. In quantum gravity, the Euclidean approach is less
well-founded because, among other problems, the action is unbounded
below.  Nevertheless, straightforward applications in
quantum gravity have tended to yield sensible answers.  We will
simply assume that this is the case for the path integral of chiral
gravity. At the end of this paper we will discuss various ways in which this assumption might fail.

\subsection{Classical saddle points}

Our first task is to determine which classical saddle points
contribute to the partition function \rref{Zsum}. These saddles
$g_c$ are solutions to the classical equations of motion with $T^2$
conformal boundary. In Euclidean signature, the bulk action of
chiral gravity is \be\label{eaction} I_{TMG} = {1\over 16 \pi G} \int d^3 x
\sqrt{g} \left(R + {2\over \ell^2}\right) + {i \ell} \int d^3
x\sqrt{g} \varepsilon^ {\lambda\mu\nu}\Gamma^{r}_{\lambda
\sigma}\left(\half\partial_{\mu}\Gamma^ \sigma_{{r}
\nu}+\frac{1}{3}\Gamma^\sigma_{\mu\tau}\Gamma^\tau_{\nu {r}} \right)
\ee The factor of $i$ appearing in front of the final term term is
the usual one that appears for Chern-Simons theories in Euclidean
signature.  To see that it must be there, note that the Chern-Simons
Lagrangian is a pseudo-scalar rather than a scalar.  Hence in
Lorentzian signature the Chern-Simons term is odd under time
reversal $t\to -t$.  Rewriting in terms of the Euclidean time
variable $ t_E= i t$ we see that the Chern-Simons action is pure
imaginary in Euclidean signature. The corresponding Euclidean
equations of motion are \be\label{eomm} {\cal G}_{\mu\nu} +{i\ell}
C_{\mu\nu}=0 \ee where ${\cal G}_{\mu\nu}$ and $C_{\mu\nu}$ are
defined as in \rref{cgdef}. One can verify directly that this is
just the Lorenztian equation of motion \rref{eom} written in terms
of a Euclidean time coordinate $t_E = it$.

The classical saddle points are smooth, real\footnote{One might wonder whether complex saddle points should be considered.  In Euclidean quantum field theory, one is instructed to include complex saddle points when, for example, momenta are held fixed at the boundary. As we are fixing the boundary metric here there is no obvious reason to include complex saddle points.} Euclidean metrics which solve \rref{eomm}. For these metrics both ${\cal G}_{\mu\nu}$ and $C_{\mu\nu}$ are real, so must vanish separately. Thus these saddle points obey the equations of motion of Einstein gravity with a negative cosmological constant
\be\label{eomreal}
{\cal G}_{\mu\nu}=0
\ee
The fact that Euclidean saddle points must be locally Einstein is in contrast with the quite difficult problem of solving the equations of motion Lorentzian signature. This  dramatic simplification will allow us to compute the partition function exactly.
One might interpret this simplification either as evidence that the Euclidean formulation does not correctly capture the complexity of the Lorentzian theory, or as evidence that the Lorentzian theory has a hidden simplicity.  Indeed it is possible that all Lorentzian solutions of chiral gravity are locally Einstein.

Solutions of the equation of motion (\ref{eomreal}) are locally isometric to three dimensional hyperbolic space $\H_3$ with Ricci curvature $R = -6/\ell^2$ .
So we just need to classify locally hyperbolic three manifolds with $T^2$ boundary.  Any locally hyperbolic three manifold is a quotient of $\H_3$ by a discrete subgroup of its isometry group $SL(2,\C)$.  In fact, it is straightforward to show (see e.g. \cite{Maloney:2007ud}) that any such smooth geometry with a $T^2$ conformal boundary must be of the form $H^3/\Z$.  We will not review this classification in detail, but simply summarize the salient points.

We will take the boundary $T^2$ to be parameterized by a complex
coordinate $z$, in terms of which the periodicity conditions are
\be\label{zident} z \sim z+1 \sim z+\tau \ee This complex coordinate
is related to the usual time and angular coordinates of global
AdS$_3$ by $z= {1\over2\pi}(\phi+it)$.\footnote{In equation (\ref{gads}) we used $\tau$ and $\phi$ to denote the time and angular coordinates of global AdS$_3$.  Here we use $t$ and $\phi$ to avoid confusion with the conformal structure parameter $\tau$.}  To find a Euclidean geometry
whose boundary has this conformal structure, write  $\H_3$ in planar
coordinates as \be\label{h3} {ds^2\over \ell^2} = {dw \bar dw + dy^2
\over y^2} \ee The conformal boundary is at $y=0$, along with the
point $y=\infty$. To obtain $\H_3/\Z$ we will quotient by the
identification \be w\sim q w,~~~~~~q=e^{2\pi i \tau} \ee If we
identify $w=e^{2\pi i z}$, then the identifications \rref{zident}
follow.  We will call the quotient $\H_3/\Z$ constructed in this way
$M_{0,1}$.

Now, the geometry described above is not the only locally hyperbolic manifold with the desired boundary behavior.  To see this, note that the geometry \rref{h3} does not treat the two topologically non-trivial cycles of the boundary $T^2$ in a democratic manner.  In particular, the $\phi$ (real $z$) cycle of the boundary torus is contractible in the interior of the geometry \rref{h3}, while the $t$ cycle is not.  In fact, for every choice of cycle in the boundary $T^2$ one can find a quotient dimensional manifold $\H_3/\Z$ which makes this cycle contractible.  A topologically nontrivial cycle $ct + d\phi$ in $T^2$ is labeled by a pair of relatively prime integers $(c,d)$.  The associated quotient $\H_3/\Z$ will be denoted $M_{c,d}$. These geometries were dubbed the ``$SL(2,\Z)$ family of black holes" by \cite{Maldacena:1998bw}.

To describe these manifolds, consider the group of modular transformations
\be\label{gammadef}
\gamma=\left({a~b\atop c~d}\right) \in SL(2,\Z)
\ee
which act as conformal transformations of the boundary $T^2$.  The cycles $t$ and $\phi$ transform as a vector $\left(t \atop \phi\right)$ under $SL(2,\Z)$, so the element $\gamma$ takes $\phi\to c\phi + dt$.  Under these transformations the conformal structure of the boundary $T^2$ is invariant, and $\tau$ transforms in the usual way
\be
\tau \to \gamma \tau = {a\tau + b \over c\tau + d}
\ee

These conformal transformations of the boundary $T^2$ extend to
isometries in the interior.  These isometries are easiest to write
down by combining the bulk coordinates $(w,y)$ into a single
quaternionic coordinate $h=w+jy$.  The modular transformation acts
as \be h\to \gamma h = (ah+b)(ch+d)^{-1} \ee Applying this isometry
to the geometry $M_{0,1}$ described in \rref{h3} gives a geometry
$M_{c,d}$ in which the cycle $c\phi+dt$ is contractible. This
geometry will represent a saddle point contribution to the partition
function. Moreover, it is possible to demonstrate the $M_{c,d}$ so
constructed are in fact the only smooth real saddle point
contributions to the partition sum.

We should emphasize that the pair of relatively prime integers $(c,d)$ determines the geometry $M_{c,d}$ uniquely.  Note that $(c,d)$ does not determine $\gamma= \left({a~b\atop c~d}\right)$ uniquely, as $a$ and $b$ are determined only up to an overall shift $(a,b)\to(a+n c,b+n d)$ for some $n\in \Z$.  However, one can check that the geometry $M_{c,d}$ is in fact independent of $n$ up to a diffeomorphism which leaves the boundary invariant.  Thus the geometries $M_{c,d}$ are in one to one correspondence with elements of the coset $SL(2,\Z)/\Z$.

We conclude that the partition function takes the form
\be
Z(\tau) = \sum_{(c,d)} Z_{c,d} (\tau)
\ee
where $Z_{c,d}(\tau)$ denotes the contribution from the saddle $M_{c,d}$.  Since the geometries are related by modular transformations we may write this as
\be
Z(\tau) = \sum_{(c,d)} Z_{0,1} (\gamma\tau)
\ee
where $\gamma$ is given by \rref{gammadef}.
The sum over $(c,d)$ may be thought of as a sum over the coset $SL(2,\Z) / \Z$. Such sums are known as Poincar\'e series and first appeared in the context of three dimensional gravity in \cite{Dijkgraaf:2000fq}.

\subsection{Sum over geometries}

We now wish to compute the perturbative partition function $Z_{0,1}(\tau)$ around the saddle point geometry $M_{0,1}$ given in \rref{h3}. The computation of the classical piece, including the gravitational Chern-Simons term, was given in \cite{Kraus:2006nb}.  This computation is rather subtle as the appropriate boundary terms must be included in the action.  The result is
\be\label{Sclassical}
e^{-I[M_{0,1}]} = q^{-k}
\ee
We note that this answer is complex, since the original Euclidean action \rref{eaction} was complex.
In particular, equation \rref{Sclassical} is holomorphic in $\tau$. It is crucial that we are at the chiral point $\mu\ell=1$, otherwise the action would not be holomorphic in $\tau$.

In order to determine the perturbative corrections to this saddle point action, we will follow the argument of \cite{Maloney:2007ud}.  The geometry $M_{0,1}$ is simply the Euclidean geometry found by imposing the identifications \rref{zident} on the global $t$ and $\phi$ coordinates of AdS$_3$.  It is therefore the usual Euclidean geometry associated with the canonical ensemble partition function computed in a fixed Anti-de Sitter background.  The partition function $Z_{0,1}$ therefore has the interpretation in Lorenztian signature as
\be\label{Zdef2}
Z_{0,1} = Tr_{\cal H} q^{L_0} {\bar q}^{\bar L_0}
\ee
where the trace is over the Hilbert space ${\cal H}$ of quantum fluctuations around a fixed Anti-de Sitter background.  The classical contribution \rref{Sclassical} may be interpreted as the contribution to this trace from a ground state $|0\rangle$ of conformal dimension $L_0 |0\rangle= -k |0\rangle$.  This ground state is just empty Anti-de Sitter space in the absence of any excitations.

At the linearized level, as shown above, the Hilbert space ${\cal H}$ includes only right moving boundary gravitons.  The linearized metrics of these boundary gravitons are obtained by acting with a right moving Virasoro generator on the empty AdS$_3$ vacuum state.
The generator $L_{-1}$ annihilates the vacuum, as $L_{-1}$ is an isometry of AdS$_3$.  The other $L_{-n}$, with $n\ge 2$ act as creation operators, and describe non-trivial boundary graviton states.  In the CFT language, such a boundary graviton is thought of as a state of the form
\be\label{lrep}
L_{-i_1}\dots L_{-i_n} |0 \rangle,~~~~~~i_n \ge 2
\ee
The resulting trace over these states is easy to compute.  It is a character of the Virasoro algebra
\be\label{z1ans}
Z_{1,0} = q^{-k} \prod_{n=2}^\infty {1\over 1-q^n}
\ee
which is closely related to the Dedekind eta function.

It is illustrative to compare this formula to equation \rref{Zsum}.  The trace over Virasoro descendants can be interpreted as the one loop contribution to the free energy; this is to be expected, as the boundary gravitons are solutions to the linearized equations of motion.  It would be interesting to derive this result directly by computing an appropriate one loop determinant, as in \cite{Giombi:2008vd}.

We now ask to what extent the formula \rref{z1ans} may be altered by higher order corrections in powers of  the inverse coupling $k^{-1}$, i.e. by the terms $I^{(n)}[g_c]$ for $n\ge 2$ in \rref{Zsum}.  We first note that the dimensions of the states appearing in the representation \rref{lrep} are completely fixed by the Virasoro algebra.  Once the dimension of the vacuum state is known, the result \rref{z1ans} is the only answer consistent with the existence of a Virasoro algebra.  This implies that equation \rref{z1ans} is one-loop exact, in the sense that the energy levels of the known states can not be altered in perturbation theory. The only possible modification of this formula at higher orders in perturbation theory in $1/k$ is a shift in the dimension of the vacuum state.  This shift is interpreted as a renormalization of the cosmological constant.  It may be absorbed by a shift of the bare coupling constant order by order in perturbation theory.

One might wonder whether there are additional states which are not present at linear order which might contribute to the sum.  We do not claim to have a complete understanding of the Lorentzian spectrum  at the non-linear level, and so can not rule out this possibility. If such states do exist, they are not solutions of the Einstein equation and hence do not appear in the Euclidean formulation followed here.  So, if the Euclidean methodology assumed here is correct, it implies that either no such additional states/corrections appear or they cancel among themselves.  In this case equation \rref{z1ans} includes the contributions to the saddle point action to all orders in the perturbation expansion in $1/k$.

Putting together the results of the previous analyses, we conclude that the partition function of chiral gravity takes the form
\be\label{Zsumm}
Z(\tau) = \sum_{c,d} Z_{0,1}(\gamma \tau),~~~~~~~Z_{0,1}(\tau) = q^{-k} \prod_{n=2}^\infty \frac{1}{1-q^n}
\ee
This sum is naively divergent, but has a well defined regularization (analogous to zeta function regularization) which is consistent with modular invariance.  We will not review the details of this regularization, which has been discussed elsewhere \cite{Maloney:2007ud,Dijkgraaf:2000fq, Manschot:2007zb, Yin:2007gv, Manschot:2007ha}, but simply state the result.

To start, we expand $Z_{0,1}(\tau)$ in powers of $q$
\be
Z_{0,1}(\tau) = \sum_{\Delta=-k}^\infty a(\Delta) q^\Delta,~~~~~~a(\Delta)=p(\Delta+k) - p(\Delta+k-1)
\ee
where $p(N)$ is the number of partitions of the integer $N$.  Then the regularization of the sum
\rref{Zsumm} is
\be
Z(\tau) = \sum_{\Delta'=-k}^0 a(\Delta') T_{-\Delta'} J
\ee
where $T_n J$ denotes the action of the $n^{th}$ Hecke operator on the modular function $J(\tau)$.  From a practical point of view, $T_n J$ may be defined as the unique holomorphic, modular invariant function on the upper half plane which has a pole of order $n$ at $\tau=i\infty$.  In particular, it is the unique $SL(2,\Z)$ invariant function whose Taylor expansion in integer powers of $q$ is
\be
T_n J(\tau) = q^{-n} + {\cal O}(q)
\ee
The coefficients in this Taylor expansion are positive integers which are straightforward to compute; we refer the reader to \cite{apostol} for a more detailed discussion of these Hecke operators and their properties.

\subsection{Physical interpretation}

The above analysis implies that, with the assumptions noted above,
the partition function of chiral gravity takes the form
\be\label{Zfinal} Z(\tau) = \sum_{\Delta=-k}^\infty N(\Delta)
q^\Delta \ee where the $N(\Delta)$ are positive integers.  These
positive integers may be computed for any desired value of $k$, as
described in detail in \cite{Witten:2007kt}. In fact, the partition
function \rref{Zfinal} is precisely the holomorphic part of the
partition function conjectured to be dual to pure gravity in
\cite{Witten:2007kt}. This is not a coincidence, as chiral gravity
apparently is a theory with all the properties shown in
\cite{Witten:2007kt} to lead uniquely to \rref{Zfinal}.  As this
partition function contains as few states as possible consistent
with modular invariance, it is referred to as the extremal CFT
partition function.

This partition function is exactly of the form that one expects for a consistent quantum theory; it is a discrete sum over a positive spectrum, with positive integer coefficients.  We contrast the present situation with that of pure Einstein gravity \cite{Maloney:2007ud}, where the corresponding computation did not yield a consistent quantum mechanical partition function unless complexified geometries were included in the sum.  The inclusion of the gravitational Chern-Simons term has resolved this apparent inconsistency.

The partition function \rref{Zfinal} has several additional interesting properties.  First, we note that the partition function makes sense only when $k$ is an integer.   Thus the cosmological constant and the Chern-Simons coefficient are quantized in Planck units.  Moreover, the spectrum of dimensions is quantized, $\Delta\in \Z$.  Thus the masses and angular momenta of all states in the theory -- including black holes --  are quantized as well.

These two rather remarkable statements are consequences of the fact that the theory is chiral.  To see this, note that in a chiral theory the partition function $Z(\tau)$ must depend holomorphically on $\tau$.   The complex structure of the boundary $T^2$ is modular invariant, so we may think of $Z(\tau)$ as a holomorphic function on the modular domain $\H_2/SL(2,\Z)$.  Including the point at $\tau=i\infty$ this modular domain may be thought of as a Riemann surface of genus zero, which is mapped analytically to the usual  Riemann sphere $\C \cup \{\infty\}$ by the j-invariant $J(\tau)$ (see e.g. \cite{apostol}). Since the partition function $Z(\tau)$ is meromorphic, it is therefore a rational function of $J(\tau)$. Moreover, if we assume that the canonical ensemble partition sum is convergent, $Z(\tau)$ must be holomorphic at all points, except possibly at $\tau=i\infty$.  Thus $Z(\tau)$ is a polynomial in the J-invariant
\be
Z = \sum_{n\ge0} a_n J(\tau)^n ,~~~~~~~J(\tau) = q^{-1} + 744 + 196884 q +\dots
\ee
for some real coefficients $a_n$
It follows that both the coupling constant and the spectrum of dimensions are quantized.

We should note, however, that this argument does not imply that the coefficients $N(\Delta)$ appearing in the expansion are positive integers.  This fact was crucial for a consistent quantum mechanical interpretation of the partition function.

These coefficients $N(\Delta)$ for large $\Delta$ can be interpreted as the exact degeneracies of quantum black holes in chiral gravity.  One can demonstrate the these coefficients reproduce precisely the black hole entropy, including an infinite series of corrections.  This is done by reorganizing the modular sum \rref{Zsumm} into a Rademacher expansion \cite{Maloney:2007ud, Dijkgraaf:2000fq, Manschot:2007ha}.  The computation proceeds exactly as in \cite{Maloney:2007ud}, so we refer the reader there for details.

Finally, we emphasize that it is not at all clear that conformal field theories with the spectrum described above exist.  No examples have been constructed with $k >1$.  Indeed, a potential objection to the existence of these theories at large $k$ was noted in \cite{Gaberdiel:2007ve, Gaberdiel:2008pr}.\footnote{See also \cite{Gaiotto:2008jt} for a discussion of these objections and \cite{Gaberdiel:2008xb} for a related discussion in the context of supersymmetric theories.}  Although the results of this paper do not imply the existence of such extremal CFTs, they certainly fit harmoniously with their conjectured existence.

\subsection{What could go wrong}
We have argued that \rref{Zfinal} follows from a conservative set of assumptions.  Nevertheless, our argument is not watertight. We now list some possible reasons why \rref{Zfinal} might not actually be the quantum chiral gravity partition function.
\begin{itemize}
\item
The Euclidean approach is invalid because the path integral is unbounded.
\item
Other complex saddle points are encountered in the analytic continuation from Lorentzian to Euclidean signature and must be included.
\item
There are nonperturbative Lorentzian classical solutions other than black holes which correspond to additional primaries in the CFT and non-perturbative corrections to the Euclidean saddle point approximation.
\item
Non-smooth saddle points must be included.
\end{itemize}
These various possibilities are not mutually exclusive.

\section{Chiral gravity, log gravity, extremal CFT and log extremal CFT}

We have presented several results pertaining to chiral gravity, log gravity, extremal CFT and their interconnections. Much remains to be understood.  In this concluding section we will draw lessons from what we learned and freely speculate on various possible outcomes. There are many possibilities -- we will limit ourselves below to the most pessimistic and the two most optimistic outcomes.

\subsection{Nothing makes sense}
The least interesting possibility, which cannot be excluded, is that
none of the theories under discussion are physically sensible. It
might turn out that at the classical level chiral gravity
has negative energy solutions, non-perturbative instabilities and/or
generically develops naked singularities. In this case the quantum
theory is unlikely to be well defined. If chiral gravity is not
classically sensible, log gravity -- which contains chiral gravity -- is
not likely to be well defined either. Extremal CFTs with large central
charges may simply not exist.
The main contraindicator to this possibility is that we
have so far discovered a rich and cohesive mathematical structure with
no apparent internal inconsistencies.

\subsection{Chiral gravity = extremal CFT}
An obvious and interesting possibility is that chiral gravity is
fully consistent and unitary, and has the modular invariant partition
function proposed in
\cite{Witten:2007kt}. In this case chiral gravity is holographically dual to a local extremal CFT .   There seems to be no room here
for non-Einstein Lorentzian solutions of chiral gravity because
there are no corresponding primaries in an extremal CFT. An
important indicator in favor of this scenario is that the torus
partition function, formally computed using Euclidean methods, gives
the extremal CFT partition function on the nose. In this case the genus g partition function of the extremal CFT would be simply the chiral-gravity
weighted sum over geometries with genus g boundary.
Conversely, if the extremal CFTs are constructed, we are finished:
we can simply declare them, in the spirit of \cite{Witten:2007kt},
to be the quantization of chiral gravity.

Of course  extremal CFTs have not been constructed for $k>1$.  Indeed, arguments against the existence of extremal CFTs at large $k$ were described in \cite{Gaberdiel:2007ve,Gaberdiel:2008pr}, although no proof was given.  An optimist might view the failure of these valiant efforts to produce an actual non-existence proof as indication that extremal CFTs do exist as highly exceptional mathematical objects.  A pessimist, on other other hand, might take the fact that no extremal CFTs have been constructed for $k>1$ as evidence that they do not exist. Further investigation is clearly needed.

\subsection{Chiral gravity $\in$ log gravity=log extremal CFT}
A third interesting possibility is the following.\footnote{We are
grateful to V. Schomerus for discussions on this point.}  Assume that quantum log gravity exists and has a
well-defined Hilbert space, and that there is a holographically dual CFT
which is logarithmic and not chiral. Of course, this is not of so
much interest in and of itself, as there is no shortage of
non-unitary quantum theories of gravity. However, chiral gravity
then also necessarily exists as the superselection sector in which
all left charges vanish. This superselection sector could still
itself have undesirable properties.  In particular there is no a
priori guarantee that it is modular invariant.  Since a modular
transformation is a large diffeomorphism in Euclidean space, this is
certainly a desirable property.  Modular invariance should be
violated if the chiral states are in some sense incomplete. For
example, consider the truncation of a generic non-chiral CFT to the
purely right-chiral sector. Generically, the only chiral operators
are the descendants of the identity created with the right moving
stress-tensor. The partition function is simply a Virasoro character
and is not modular invariant.  In the previous section it was argued that in the context of chiral gravity the primaries associated to black holes complete, in the manner described in \cite{Witten:2007kt}, this character to a modular invariant
partition function.  However, it is possible that no such completion exists.
It might be that chiral gravity is in some sense a physical, unitary  subsector of log gravity, but its dual does not obey all the axioms of a local CFT.  Interestingly, in \cite{bilal} it was found that some 2D gravity theories coupled to matter are  logarithmic CFTs.

At first the compelling observation that the Euclidean computation
of the chiral gravity partition function gives the extremal CFT
partition function would seem to be evidence against
this possibility. One would expect that any extra states present in log gravity
would spoil this nice result. However, as log
gravity is not unitary, the extra contributions to the partition
function can vanish or cancel. Indeed it is a common occurrence in
logarithmic CFTs for the torus partition function to contain no
contributions from the logarithmic partners. We see hints of this
here: as $c_L=0$, the left-moving gravitons of log gravity have zero
norm and hence do not contribute. This suggests the at-present-imprecise notion of a
``log extremal CFT": a logarithmic CFT whose partition
function is precisely the known extremal partition function.
Perhaps previous attempts to construct extremal CFTs have failed precisely because the theory was assumed to be unitary rather than logarithmic. Clearly there is much to be understood and many interesting avenues to pursue.

 \section*{Acknowledgements}
This work was partially funded by DOE grant DE-FG02-91ER40654. The
work of W.~S. was also funded by NSFC.  The work of A. M. is
supported by the National Science and Engineering Research Council
of Canada.  We wish to thank S. Carlip, S. Deser, T. Hartman, M. Gaberdiel, S. Giombi, G. Giribet,  D.  Grumiller, C. Keller,
M. Kleban, G. Moore, M. Porrati, O. Saremi, V. Schomerus, E. Witten, A. Wissanji and X. Yin
for useful conversations and correspondence. W. S. thanks the High Energy Group at
Harvard and McGill Physics Department for their kind hospitality.


\end{document}